%% file: MAIN_EMSE.tex
\RequirePackage{fix-cm}
\documentclass[smallextended]{svjour3}       %
\smartqed  %

\usepackage{cite}
\usepackage{amsmath,amssymb,amsfonts}
\usepackage{algorithmic}
\usepackage{graphicx}
\usepackage{textcomp}
\usepackage[table]{xcolor}
\usepackage{booktabs}
\def\BibTeX{{\rm B\kern-.05em{\sc i\kern-.025em b}\kern-.08em
    T\kern-.1667em\lower.7ex\hbox{E}\kern-.125emX}}

\usepackage[backgroundcolor=white,bordercolor=blue,linecolor=blue]{todonotes}
\usepackage{xspace}
\usepackage{blindtext}
\usepackage{hyperref}

\usepackage{multirow}
\usepackage{makecell}
\usepackage{subcaption}
\usepackage{setspace}

\usepackage{balance}
\usepackage{array}
\usepackage{lipsum}
\usepackage{listings}
\usepackage{tcolorbox}

\usepackage[framemethod=TikZ]{mdframed}
\usepackage{framed}
\usepackage{tikz}
\usepackage{lipsum}

\usepackage{marginnote}
\usepackage[ruled,vlined,lined,commentsnumbered]{algorithm2e}
\usepackage{marvosym}
\usepackage{orcidlink}

\input{cmd}

\begin{document}

\title{Fuzzing-based Mutation Testing of C/C++ Software in Cyber-Physical Systems}

\author{
Jaekwon Lee 
\orcidlink{0000-0002-5222-1820}
\and %
Fabrizio Pastore 
\orcidlink{0000-0003-3541-3641}
\and  %
Lionel Briand %
\orcidlink{0000-0002-1393-1010}
}

\institute{
            Jaekwon Lee \at
            {Kangwon National University, Chun-cheon, KR} \\\Letter~\email{jaekwon.lee@kangwon.ac.kr} \\
        \and
            Fabrizio Pastore \at 
            {SnT, University of Luxembourg, Luxembourg, LU} \\
            \email{fabrizio.pastore@uni.lu} \\
        \and
            Lionel C. Briand \at
            {Lero, University of Limerick,  Limerick, IE} \\
            {University of Ottawa, Ottawa, CA} \\
            \email{lbriand@uottawa.ca} \\
        \and
            This work was partially done while J. Lee was affiliated with the University of Ottawa and the University of Luxembourg. \\
}

\date{Received: date / Accepted: date}

\maketitle

\sloppy   %

\begin{abstract}
Mutation testing can help minimize the delivery of faulty software. Therefore, it is a recommended practice for developing embedded software in safety-critical cyber-physical systems (CPS). However, state-of-the-art mutation testing techniques for C and C++ software, which are common languages for CPS, depend on symbolic execution. Unfortunately, symbolic execution's limitations hinder its applicability (e.g., systems with black-box components).

We propose relying on fuzz testing, which has demonstrated its effectiveness for C and C++ software. Fuzz testing tools automatically create test inputs that explore program branches in various ways, exercising statements in different program states, and thus enabling the detection of mutants, which is our objective.

We empirically evaluated our approach using software components from operational satellite systems. Our assessment shows that our approach can detect between 40\% and 90\% of the mutants not detected by developers' test suites. Further, we empirically determined that the best results are obtained by integrating the Clang compiler, a memory address sanitizer, and relying on laf-intel instrumentation to collect coverage and guide fuzzing. Our approach detects a significantly higher percentage of live mutants compared to symbolic execution, with an increase of up to 50 percentage points; further, we observed that although the combination of fuzzing and symbolic execution leads to additional mutants being killed, the benefits are minimal (a gain of less than one percentage point).
\end{abstract}


\input{introduction}

\input{related}

\input{approach}

\input{approach_cpp}

\input{evaluation-new}

\input{conclusion}

\input{ack}

\bibliographystyle{IEEEtran}
\bibliography{./bibliography/ref}

\end{document}

%% file: cmd.tex
\newcommand{\APPR}{\emph{MOTIF}\xspace}
\newcommand{\MOTIFC}{\emph{MOTIF-C}\xspace}
\newcommand{\MOTIFCPP}{\emph{MOTIF-C++}\xspace}
\newcommand{\SEMU}{\emph{SEMu}\xspace}
\newcommand{\SEMUp}{\emph{SEMuP}\xspace}
\newcommand{\MOTIF}{\APPR}

\newcommand{\SAIL}{\emph{ESAIL}\xspace}
\newcommand{\Sentinel}{\emph{S5}\xspace}

\newcommand{\ASNLib}{\emph{ASN1lib}\xspace}

\newcommand{\UTIL}{\emph{LIBU}\xspace}

\newcommand{\MLFS}{\emph{MLFS}\xspace}

\newcommand{\CR}[2]{{#2}}

\newcommand{\ENDUPDATED}{\color{black}}

\definecolor{mGreen}{rgb}{0,0.6,0}
\definecolor{mGray}{rgb}{0.5,0.5,0.5}
\definecolor{mPurple}{rgb}{0.58,0,0.82}
\definecolor{backgroundColour}{rgb}{0.95,0.95,0.92}

\lstdefinestyle{CStyle}{
    backgroundcolor=\color{white},   
    commentstyle=\color{mGreen},
    keywordstyle=\color{magenta},
    numberstyle=\small\color{mGray},
    stringstyle=\color{mPurple},
    basicstyle=\small\ttfamily,
    breakatwhitespace=false,         
    breaklines=true,                 
    captionpos=b,                    
    keepspaces=true,                 
    numbers=left,                    
    numbersep=1em,                  
    showspaces=false,                
    showstringspaces=false,
    showtabs=false,                  
    tabsize=2,
    language=C,
    xleftmargin=2.5em,
    xrightmargin=0.5em,
    framexleftmargin=2em,      %
    frame=tb,
    linewidth=\textwidth
}

\lstdefinestyle{CPPStyle}{
    backgroundcolor=\color{white},   
    commentstyle=\color{mGreen},
    keywordstyle=\color{magenta},
    numberstyle=\small\color{mGray},
    stringstyle=\color{mPurple},
    basicstyle=\small\ttfamily,
    breakatwhitespace=false,         
    breaklines=true,                 
    captionpos=b,                    
    keepspaces=true,                 
    numbers=left,                    
    numbersep=1em,                  
    showspaces=false,                
    showstringspaces=false,
    showtabs=false,                  
    tabsize=1,
    language=C++,
    xleftmargin=2.5em,
    xrightmargin=0.5em,
    framexleftmargin=2em,      %
    frame=tb,
    linewidth=\textwidth,
}

\definecolor{light-gray}{gray}{0.9}

\mdfsetup{skipabove=5pt,skipbelow=3pt}
\mdfdefinestyle{RQFrame}{%
	linecolor=black,
	outerlinewidth=0.15pt,
	roundcorner=3pt,
	innertopmargin=2pt,
	innerbottommargin=2pt,
	innerrightmargin=4pt,
	innerleftmargin=4pt,
	backgroundcolor=light-gray,
	nobreak=true,
	}

\newboolean{showcomments}
\setboolean{showcomments}{true}
\ifthenelse{\boolean{showcomments}}
{   
    
}
{
    
}

\definecolor{yellow}{RGB}{255,255,153}
\definecolor{grey}{RGB}{224,224,224}
\definecolor{DarkOrange}{rgb}{0.8,0.3,0.0}

\newcommand{\RESUB}[2]{#2}
\newcommand{\RESUBSTART}{\color{black}}
\newcommand{\RESUBEND}{\color{black}}

\newcommand{\REV}[2]{\textcolor{black}{#2}}
\newcommand{\REVSTART}{\color{black}}
\newcommand{\REVEND}{\color{black}}

%% file: introduction.tex
\section{Introduction}
\label{sec:introduction}
Ensuring high-quality test suites is essential for quality assurance of embedded software in cyber-physical systems (CPS); indeed, this is the reason why independent software validation and verification activities are required by safety standards~\cite{ecss40C}. \RESUB{R1.2}{Further, although embedded software for CPS can be developed using dedicated visual languages~\cite{Buffoni2021}, our experience in the space context shows that most software is directly implemented using C and C++, which suggests that approaches supporting validation and verification shall target those languages.}

Mutation analysis is an effective method to evaluate the quality of a test suite. It involves measuring the mutation score, which is the proportion of programs with artificially injected faults (mutants) detected by the test suite~\cite{papadakis2019mutation}. There is a strong association between a high mutation score and a high fault-revealing capability for test suites~\cite{papadakis2018mutation,Chekam:17}. Additionally, recent studies have demonstrated that mutation analysis can be applied cost-effectively to large CPS software~\cite{Oscar:TSE}.

In practice, mutation analysis warrants the selection of inputs for mutation testing, as test cases should ideally detect all or at least a significant proportion of the generated mutants. A mutant detected by a test suite is considered killed. However, due to the typically high number of mutants generated in large CPS projects, it is challenging for engineers to perform mutation testing manually.

Unfortunately, we lack automated test data generation techniques (\emph{automated mutation testing} techniques) suitable for the mutation testing of CPS. Indeed, most existing techniques do not target C and C++, which are prevalent in CPS. Further, the state-of-the-art solution for automated mutation testing of C software, \SEMU~\cite{Chekam2021}, is based on the KLEE symbolic execution engine~\cite{KLEE}. While effective for command line utilities, it inherits the limitations of symbolic execution. Specifically, it requires environmental modeling (e.g., network communication) and cannot be applied to programs needing complex analyses for input generation (e.g., programs with floating point instructions). Additionally, it generates test inputs for command line utilities, which are rarely used in CPS, and does not produce unit test cases or target other CPS interfaces.

Search-based techniques developed for other programming languages (e.g., Java~\cite{fraser2011mutation}) are impractical for C and C++ software due to the difficulty of instrumenting the software to compute dedicated fitness functions (e.g., branch distance). For instance, computing branch distance at runtime necessitates modifying all conditional statements in the software under test (SUT), requiring the source code to be processed with static analysis tools that load all dependencies.  Configuring these tools to process the multiple source files in large systems is often impractical unless the tool is well integrated with the compiler used for the SUT. Moreover, CPS source files often rely on architecture-specific C constructs (e.g., for the RTEMS compiler~\cite{RTEMS}) that are not successfully parsed by static analysis frameworks~\cite{Oscar:TSE}.

Different from search-based testing techniques,  grey-box fuzz testing techniques~\cite{manes2019art} (hereafter, \emph{fuzzers}) can generate test data without relying on complex code instrumentation and thus can be easily applied to C and C++ software. However, fuzzers target console software, while CPS software is usually tested either with system-level test scripts interacting with a hardware emulator or through unit and integration test cases implemented with the same language as the SUT. In this paper, we focus on the automated generation of unit test cases because fuzzing large systems is an open research problem~\cite{KimXSO22}. 
Approaches for fuzzing function calls at unit or API level exist~\cite{FuzzGen,FUDGE,APICRAFT,Jeong2023,Hopper}, although in the case of large CPS software, their applicability might be limited by static analysis' scalability issues and difficulties in processing embedded libraries. Further, approaches to generate fuzzing drivers for mutation testing do not exist.

We propose \emph{MutatiOn TestIng with Fuzzing} (\MOTIF), an approach that automatically generates fuzz drivers, leverage the data produced by fuzzers, and exercise a mutated function and the corresponding original function, looking for diverging outputs to detect mutants. \MOTIF automatically generates seed files for fuzzers and integrates strategies to process mutant-killing inputs to eliminate false positives due to nondeterminism\REV{C1.1}{; in this context, possible sources of nondeterminism are functions interacting with timers, accessing the file system, or receiving inputs from sensors}.

Instead of designing a dedicated fuzzing algorithm, \MOTIF leverages state-of-the-art (SOTA) fuzzers\REV{C1.2}{, as their bucket-based coverage strategy has demonstrated being effective in maximizing the exploration of different execution paths. %
}%
This strategy not only selects inputs that lead to different software states\REV{C1.2}{, which could help killing mutants,} but enables tracking the behavioral differences between the original and mutated functions \REV{C1.2}{since such differences lead to distinct execution paths that can be exercised only by mutant-killing inputs.} \RESUB{R1.2}{Although \MOTIF can target any C/C++ program, we designed it taking into account characteristics that are prominent in embedded software for CPS. Specifically, CPS software is not executed from the textual console but directly within the operating system, largely relies on floating point instructions, and leverages specific OS features (e.g., RTEMS) and libraries that complicate its analysis or symbolic interpretation. For these reasons, we present \MOTIF as a tool for C/C++ embedded software for CPS.}

We introduced \MOTIF in a paper presented at the 2023 International Conference on Automated Software Engineering~\cite{MOTIF} where we reported on its effectiveness on three case study subjects in the space domain developed using the C language and demonstrated that it outperforms mutation testing based on symbolic execution. In this paper, we extend our previous work by:
\RESUBSTART{}
\begin{itemize}

\item Performing a large empirical study to investigate (RQ1) the best fuzzing configurations (compiler, sanitizers, and coverage metric) for mutation testing. Note that the identification of the best fuzzing configuration for different application contexts remains an open problem~\cite{EURECOM7801}. 
\REV{C2.1}{We thus focused on identifying the best configurations for safety-critical space software implemented in C and C++, including mathematical libraries, control software, and data processing software; our results shall thus generalize to other types of CPS including these kinds of components, for example, avionic and robotic CPS.}
Our experiments show that the best results are obtained when 
relying (1) on the \texttt{Clang} compiler, which leads to the quickest fuzz drivers, (2) the address sanitizer (\texttt{ASAN}), which prevents false positives due to violation of function preconditions, and (3) the \emph{LAF-Intel}~\cite{LAF:Blog} optimization for coverage, \RESUB{R2.4}{which enables computing a better fitness score for mathematical functions, which are, in turn, among the most critical software components.} 
Specifically, the best fuzzing configuration, compared to a standard fuzzing configuration (i.e., AFL++~\cite{aflchurn} with the GCC~\cite{GCC} compiler),  leads to an improvement varying between 1.53 and 4.34 percentage points. Although limited, such improvements may still enable the detection of critical faults; for example, when reusing the generated test cases for regression testing. Further, our findings help engineers make informed decisions on projects setups. For example, mutation testing improvements may not justify a large development endeavor to enable a project to be recompiled with Clang. \RESUB{R1.3}{However, when compiling projects with Clang, ASAN, and LAF-Intel is feasible (e.g., it is sufficient to change a few configuration parameters), the best configuration is a better choice (e.g., enables testing additional boundary cases) and might be considered also for other fuzzing applications beyond mutation testing}.

\item Assessing (RQ4) the performance gain achieved by relying on hybrid-fuzzing, which combines fuzzing and symbolic execution. Our results confirm the findings of related work that demonstrate that hybrid fuzzing leads to improved test effectiveness; 
however, in our experiments, its applicability is limited as it helps significantly \REV{2.7}{improve the mutation score for only one in four subjects}. 
The key advantage observed with hybrid-fuzzing is the reduction of \REV{C1.3}{the median time required to kill a mutant in three out of four subjects (between 14\% and 32\% faster),} which suggests using hybrid-fuzzing for large projects leading to a large number of mutants. 
\item Evaluating (RQ5) the cost-effectiveness of reusing mutant-killing inputs to test live mutants; our results suggest that reusing mutant-killing inputs to kill other
mutants is recommended when there is a sufficient test budget
of at least 5000 seconds per mutant \REV{C2.2}{in our subjects and, possibly, in similar systems}.
\item Replicating, with the best fuzzing configuration, our original~\cite{MOTIF} assessments of the complementarity between fuzzing and symbolic execution for mutation testing (RQ2), and  the contribution of \MOTIF's seeding strategy to its results (RQ3). With our subjects, \MOTIF kills between 39.5\% and 88.7\% of mutants, on average, with most of the mutants (52\% to 95\%)  being killed by the search process, not the seed inputs. \RESUB{R1.1}{Further, \MOTIF outperforms, by 13.18 to 50.33 percentage points, the results obtained with the mutation testing process based on symbolic execution implemented by the state-of-the-art tool \SEMU, thus suggesting that \MOTIF is the most appropriate solution for the mutation testing of CPS embedded software. \RESUB{R2.4}{This is the first work comparing fuzzing and symbolic execution for mutation testing.}}

\item Contributing to the software engineering literature with the first study of mutation testing on industrial software for CPS.
In this paper, we assess \MOTIF on the control software of ESAIL\cite{ESAILESA}, a satellite currently on orbit, and \emph{Sentinel-5 UVNS L1b Prototype Processor}, a ground software processing radiation data collected by the Sentinel mission of the European Space Agency (ESA)~\cite{esaSentinel}, in addition to the subjects considered in our conference paper: a mathematical library qualified for flight systems, a commercial utility library for nanosatellites,  and a serialization/deserialization library, all made available in the context of a project with ESA~\cite{FAQAS}. 

\item Providing a technical solution enabling the application of \MOTIF to object-oriented C++ programs. Since state-of-the-art mutation testing tools do not target C++ software, our solution enables mutation testing assessment on a broader set of case study subjects.

\end{itemize}

\MOTIF is available online~\cite{MOTIFGIT}; also, we provide a \emph{replication package} with our open-source subjects and all our empirical data~\cite{REPLICABILITY}.

\RESUBEND{}

The paper proceeds as follows. Section~\ref{sec:related} provides background (symbolic execution and fuzzing) and related work (automated mutation testing). Section~\ref{sec:approach} describes \MOTIF.
Section~\ref{sec:approach:cpp} presents our extensions to deal with object-oriented programs in C++. Section~\ref{sec:evaluation} presents our empirical evaluation. Section~\ref{sec:conclusion} concludes the paper.

%% file: related.tex
\section{Background and Related Work}
\label{sec:related}

This paper relates to the enormous body of work on automated mutation testing and fuzzing; selected, relevant work is discussed below.

\subsection{Symbolic execution}
\label{sec:symbex}

Symbolic execution (SE) is a program analysis technique that relies on an interpreter to process the source code of the SUT and automatically generate test inputs~\cite{Anand2013}. 
Inputs are represented through symbolic values; during the symbolic execution,  the state of the SUT includes the symbolic values of program variables at that execution point, a path constraint on the symbolic values to reach that point, and a program counter. 
The path constraint is a boolean formula that captures the conditions that the inputs must satisfy to follow that path. Constraint solving~\cite{SATandCPsurvey:2006} is then used to identify assignments for the symbolic inputs that  satisfy the path constraint.

SE presents several limitations, including (1) the need for abstract representations  for the external environment and any black-box components used by the SUT---otherwise, the SE engine cannot know what outputs to expect from the environment, (2) path explosion---the SE engine may need to process a large number of paths before satisfying a target predicate, (3) path divergence---abstract representations do not behave like the real systems, (4) the handling of complex constraints, e.g., solving constraints with floating point variables.

A recent solution to partially address the above-mentioned limitations is 
 dynamic symbolic execution (DSE), which consists of treating only a portion of the program state symbolically. Concrete program states help dealing with complex constraints or path explosion (e.g., SE is used after a certain branch has already been reached using a concrete input).
 However, most frameworks with DSE capabilities, e.g., Angr~\cite{shoshitaishvili2016state}, KLEE~\cite{KLEE}, and S2E~\cite{S2E}, %
 require some degree of environment modeling (e.g., libc library modeling in KLEE), which limits their practical applicability~\cite{SymCC}.

Compilation-based approaches like QSYM~\cite{yun2018qsym}, SYMCC~\cite{SymCC}, SymQEMU~\cite{poeplau2021symqemu}, and SYMSAN~\cite{SYMSAN} augment the original program with instructions to populate and solve symbolic expressions while the original software is executed; such characteristic eliminates some limitations of interpretation-based approaches,  thus being applicable to a broader set of software systems. For example, since the symbolic execution interacts with the actual environment there is no need to emulate it within the interpretation layer. SYMCC requires the source code of the SUT, while QSYM relies on dynamic binary instrumentation.
SymQEMU, instead, extends the applicability of SYMCC to binary programs by relying on the QEMU emulator for code instrumentation, while SYMSAN relies on dynamic data-flow analysis to reduce the cost of symbolic state management.
Since the above-mentioned symbolic execution approaches have shown to provide their best results when combined with fuzzing, a solution referred to as \emph{hybrid fuzzing} (see Section~\ref{sec:background:fuzzing}), we considered them when assessing the integration of \MOTIF with hybrid fuzzing (see Section~\ref{sec:evaluation}).

\subsection{Fuzzing}
\label{sec:background:fuzzing}

Fuzzing (or fuzz testing) is an automated testing technique 
that generates test inputs by repeatedly modifying\footnote{To avoid confusion, we avoid the term `mutation' when describing fuzzing techniques.} existing inputs; the selection of the inputs to modify is usually driven by metrics collected during the execution of the SUT.
Depending on the information collected during program execution, fuzzing techniques (i.e., fuzzers) are classified as black-box, white-box, or gray-box. 

In this paper, we focus on \emph{grey-box fuzzers} because they have demonstrated to effectively maximize code coverage~\cite{FuzzBenchPaper} and discover faults~\cite{ICST:22:Sarro} (mainly crashes and memory errors), two objectives that relate to the problem studied in this paper; indeed, to kill a mutant it is necessary to (1) exercise a mutated statement, which can be achieved by maximizing code coverage, and (2) exercise the mutated statement with many different inputs (i.e., in different states), a common practice in fuzzers to discover crashes and memory errors.

Most fuzzers generate input files to be used for system-level testing of console applications and 
engineers are therefore required to implement driver programs (hereafter, \emph{fuzzing drivers}) that rely on the data generated by the fuzzer to test other software interfaces (e.g., APIs, see Section~\ref{sec:motif:step1}). Most fuzzers keep a pool of input files and rely on the following evolutionary search process: (1) select an input file from the pool, (2) modify the input file to generate new input files, (3) provide the new input files to the SUT and monitor its execution, (4) report crashes or problems detected through code sanitizers (hereafter, sanitizers~\cite{UBSan}), (5) add to the pool all the input files that contribute to improve code coverage.

What facilitates the adoption of fuzzers is that they rely on simple dynamic analysis strategies to trace branch coverage of C/C++ programs. A common strategy consists of dynamically identifying branches by applying a hashing function to the identifiers assigned to code blocks by compile-time instrumentation; it is implemented as an extension of popular C/C++ compilers~\cite{AFL:instr}. Further, instead of relying on traditional branch coverage~\cite{ammann2016introduction}, most fuzzers adopt a bucketing approach to track the number of times each branch has been covered by each input file: only once, twice, three times, between four and seven, between 8 and 15, between 16 and 31, etc.; the fuzzers add to the pool those files that cover a bucket not observed before for at least one branch. Such bucketing strategy helps reach software states that are not reachable by simply relying on branch coverage. 

Fuzzers mainly differ with respect to the strategy adopted to (1) select what operations to apply in order to modify input files and obtain new ones (e.g., MOpt~\cite{MOPT} relies on a particle swarm optimization algorithm) and (2) select the inputs from the input pool (e.g., AFLfast~\cite{AFLfast} and AFL++~\cite{AFL++} rely on a simulated annealing algorithm and prioritize new paths and paths exercised less frequently). Also, fuzzers differ in the strategy adopted to determine interesting inputs. For example, \emph{directed grey-box fuzzers}~\cite{WangDGF2020}, instead of maximizing code coverage, aim to reach specific targets,  usually a subset of program locations (e.g., modified code) or invalid sequences of operations (e.g., use-after-free).

 \emph{Hybrid fuzzers}~\cite{Hybrid,Stephens2016,yun2018qsym}, instead, rely on grey-box fuzzing to explore most of the execution paths of a program and leverage DSE to explore branches that are guarded by narrow-ranged constraints when the fuzzer does not improve coverage further. 
 Well-known hybrid fuzzing solutions consist of combining AFL~\cite{AFL} with QSYM~\cite{yun2018qsym} and SYMCC~\cite{SymCC}, which have been found to  outperform earlier approaches such as Driller~\cite{Stephens2016} and hybrid testing~
 \cite{Hybrid}. The SOTA approach is Fuzzolic~\cite{FUZZOLIC}, which relies on QEMU to generate symbolic queries and solves them with a fuzzing-based technique~\cite{fuzzySAT}.

Some researchers have addressed the problem of generating test drivers to fuzz test program functions as in unit testing~\cite{FuzzGen,FUDGE,APICRAFT,Jeong2023,Hopper}; however, except for Hopper~\cite{Hopper}, they make the assumption that the function under test are already integrated into consumer programs (i.e., programs using the library API) or unit tests~\cite{Jeong2023}, and all of them rely on complex static and dynamic analyses which 
are
infeasible with large CPS software. Nevertheless, although existing tools do not target the generation of drivers for mutation testing, studying their integration into \MOTIF is part of future work.

To avoid relying on program analysis, building on the potential shown by LLM-based test case generation~\cite{LLMtestingSurvey,shin2024,Fan:2023}, researchers and developers are investigating LLM-based fuzz driver generation~\cite{Zhang2024,OSS:fuzz:llm,Lyu2024}.
For example, a recent study has demonstrated the feasibility of relying on large language models (LLMs) to automatically generate fuzz driver~\cite{Zhang2024}; specifically they could generate effective drivers (compile and cover the code) for 91\% of the 86 APIs under test with, however, some limitations such as violating API protocol (e.g., parameter initialization) in 39\% of the cases. 
LLM-based generation of mutation testing drivers in \MOTIF is part of future investigations.

\CR{C.3}{Other techniques address the problem of generating highly structured input files~\cite{TensileFuzz,Skyfire}. TensileFuzz generates structured inputs (e.g., image or zip files) by probing random executions to derive constraints for potential input fields, and then relying on string constraint solving to derive inputs~\cite{TensileFuzz}. SkyFire, instead, learns a probabilistic context-sensitive grammar to generate JSON and XML files~\cite{Skyfire}. Such techniques can generate input files with a complex structure but they do not generate unit test cases, which is necessary in our context; however, leveraging those approaches to  populate complex data structures may also help with unit-level fuzz testing.}

\subsection{Automated mutation testing}
\label{sec:background:mt}
To kill a mutant, a test case should satisfy three conditions: \emph{reachability} (i.e., the test case should execute the mutated statement), \emph{necessity} (i.e., the test case should cause an incorrect intermediate state if it reaches the mutated statement), and \emph{sufficiency} (i.e., the observable state of the mutated program should differ from that of the original program)~\cite{offutt1997automatically}. 
Automated mutation testing approaches differ regarding the strategy adopted to satisfy these conditions.

There exist two families of 
automated mutation testing techniques based respectively on:
\emph{constraint solving} and \emph{meta-heuristic search}.
Only one of them relies on fuzzing~\cite{brown2020mutation}, as further described below.

In this Section, we mainly focus on techniques targeting C and C++ programs because these languages are used in many CPS; unfortunately, the C and C++ languages are more complex to process for static and dynamic analysis techniques than the higher-level languages targeted by most of the techniques in the literature (e.g., Java).

\subsubsection{Techniques based on constraint solving}
\label{sec:back:cs}

Inspired by the earlier work of Offut et al.~\cite{offutt1997automatically}, Holling et al. execute symbolically the original and mutated functions with input data leading them to generate different outputs~\cite{holling2016nequivack}. A similar technique from Riener et al.~\cite{riener2011test} relies on a bounded model checker (BMC) to select the input values that kill the mutant. 
Unfortunately, no prototype tools for the above-mentioned approaches are available.

The SOTA tool for automated mutation testing is \SEMU~\cite{Chekam2021,SEMUgit}, which relies on KLEE to generate test inputs based on SE. 
To speed up mutation testing, \SEMU relies on meta-mutants (i.e., it compiles mutated statements and the original statements together). First, \SEMU relies on SE to reach mutated statements (reachability condition).
Then, for each mutant, it relies on constraint solving to determine if inputs that weakly kill the mutant exist (necessity condition). For killable mutants, it symbolically runs the mutated and the original program in parallel; when an output statement is reached (e.g., a \texttt{printf} or the \texttt{return} statements of the main function), it relies on constraint solving to identify input values that satisfy the sufficiency condition.

\subsubsection{Techniques based on meta-heuristic search}

Most of the work on automated mutation testing with meta-heuristic search targets Java software; we report the most relevant techniques below.
Ayari et al.~\cite{ayari2007automatic} rely on an Ant Colony Optimization algorithm~\cite{dorigo2006ant}
 driven by a fitness function that focuses on the reachability condition. Precisely, their fitness measures the distance (number of basic blocks in the program's control flow graph) between the mutated statement and the closest statement reached by a test case. 
Fraser and Zeller~\cite{fraser2011mutation}, instead, extended the EvoSuite tool~\cite{EvoSuite} with a fitness function considering the reachability and the necessity conditions (number of statements that are covered a different number of times by the original and the mutated program). The integration of mutation testing into EvoSuite has been further improved with branch distance metrics tailored to the operator used to generate the mutants~\cite{fraser2015achieving}. Recently, EvoSuite has been further extended by Almulla et al. with adaptive fitness function selection (AFFS), a hyperheuristic approach that relies on reinforcement learning (RL) algorithms to determine which composition of fitness functions to use~\cite{Almulla2022}. 
Unfortunately, when applied to mutation testing, AFFS does not perform better than SOTA solutions~\cite{fraser2015achieving}.

Concerning C software, we should note the work of Souza et al.~\cite{souza2016strong}, who rely on the Hill Climbing AVM algorithm~\cite{KorelAVM}.
They combine three fitness functions that rely on branch distance to measure how far an input is from satisfying each of the three killing conditions.
The mutation score obtained with simple C programs ranges between 52\% and 93\%.
The approach has been implemented on top of AUSTIN, a search-based test generation tool for C~\cite{lakhotia2010austin,LAKHOTIA2013112,AUSTIN}; however, this implementation is not available.
A recent search-based testing tool prototype for C is Ocelot~\cite{Scalabrino:18}; however, it has not been extended for automated mutation testing. 
Another key limitation of both Ocelot and AUSTIN is that they implement preprocessing steps that do not work with complex program structures (e.g., we couldn't apply them to the subject programs considered in our empirical evaluation because of preprocessing errors).

A recent mutation testing technique targeting C software is that of Dang et al.~\cite{dang2019efficiently}, who propose a co-evolutionary algorithm that reduces the search domain at each iteration (the original search domain is replaced by the joint domain of the best solutions found); unfortunately, their prototype is not available.

\subsection{Techniques based on fuzzing}
The work of Bingham~\cite{brown2020mutation} is the only one to rely on fuzzing to automate mutation testing for C software.
For input generation, it relies on TOFU~\cite{wang2020tofu}, a grey-box, grammar-aware
fuzzer that
generates grammar-valid inputs by modifying existing ones. 
Similar to Ayari's work, TOFU's input generation strategy is guided by the distance between the mutated statement and the closest statement reached by a test case; however, instead of generating unit test cases, it generates input files matching a given grammar.
Unfortunately, the results obtained by Bingham are preliminary (they targeted only the Space benchmark~\cite{SPACE}) and a prototype tool is not available.

\RESUB{R3.2}{DifFuzz~\cite{DifFuzz}, instead, executes two distinct versions of a program and compares their coverage and execution cost (e.g., time) to identify inputs that trigger side channel attacks; although this procedure might be leveraged to perform mutation testing (i.e., execute the original program and the mutant), its implementation targets Java systems. Further it is worth noting that DifFuzz does not compute any difference in terms of code coverage, as well as, the difference in execution time 
is delegated to a driver that needs to be manually implemented by the software engineer, thus limiting the usability of the approach.}

\CR{C.4}{Mu2~\cite{MU2}, which has been developed in parallel with \MOTIF, is a fuzzer that integrates the findings of search-based unit test generation~\cite{EvoSuiteMutation} to generate test input files with fuzzing: it relies on the mutation score to drive the generation of test inputs. Different from \MOTIF, which tests each mutant independently from the others, potentially with different inputs, Mu2 tests every live mutant with each generated input and, in the input pool, prioritizes those inputs that increase the mutation score.
Results show that Mu2 kills more mutants than the inputs generated by a traditional fuzzer to test the original function. However, it is unclear whether Mu2's approach (i.e., testing all the live mutants together) is more effective than that of  \MOTIF (i.e., executing a fuzzer to test the original and mutated function in sequence). Unfortunately, a direct comparison of Mu2 and \MOTIF is not feasible because
the scalability of Mu2 is enabled by dynamic classloading and instrumentation, two options that are feasible for Java programs but not for the C/C++ programs targeted by \MOTIF. Further, by targeting Java, Mu2 can easily determine if mutants are killed by relying on the method `equals', which is implemented by every class to determine if two instances are equal; the method `equals' is not available in C and C++ software. 
Mu2's results follow previous work showing that, in Java benchmarks, prioritizing inputs that increase the mutation score may lead to higher branch coverage and mutation score than traditional prioritization strategies based on branch coverage~\cite{CoverageGuided}.}

\RESUBSTART

\subsection{Industrial applications of mutation testing}

Most mutation testing approaches target Java software; the few mutation testing approaches targeting C/C++ software have been assessed with open source console utilities~\cite{Chekam2021} or simple algorithm implementations (e.g., merge sort)~\cite{holling2016nequivack,riener2011test}, but there is no work assessing mutation testing tools on C/C++ industrial software.
Last, most empirical studies on large industrial software systems concern mutation analysis~\cite{Just2021Practical}, not mutation testing.

\RESUBEND

\subsection{Summary}
To summarize, our research is motivated by the lack of support for automated mutation testing of C/C++ software. The SOTA approach for the automated mutation testing of C/C++ software (i.e., \SEMU) relies on KLEE and inherits its limitations, making it inapplicable to most CPS software; further, it does not generate unit test cases but selects inputs for console programs. Other SE tools (QSYM and SYMCC) also present technical limitations preventing their application to CPS software. Search-based approaches for the mutation testing of C/C++ software present acute feasibility challenges due to static analysis, which is needed for branch distance fitness but does not scale in large software projects. Though fuzzing appears to be a feasible input generation strategy for mutation testing, existing fuzzers do not generate test drivers for unit testing. The only fuzzer proposed for mutation testing is not available for download and its results are very preliminary.

\RESUBSTART
Last, our paper addresses the lack of empirical assessments of mutation testing on industrial software by targeting libraries and control software used in satellite systems as case studies.
\RESUBEND

%% file: approach.tex
\section{Proposed approach: MOTIF}
\label{sec:approach}

\label{sec:motif}
\begin{figure}[tb]
\begin{center}
\includegraphics[width=0.9\linewidth]{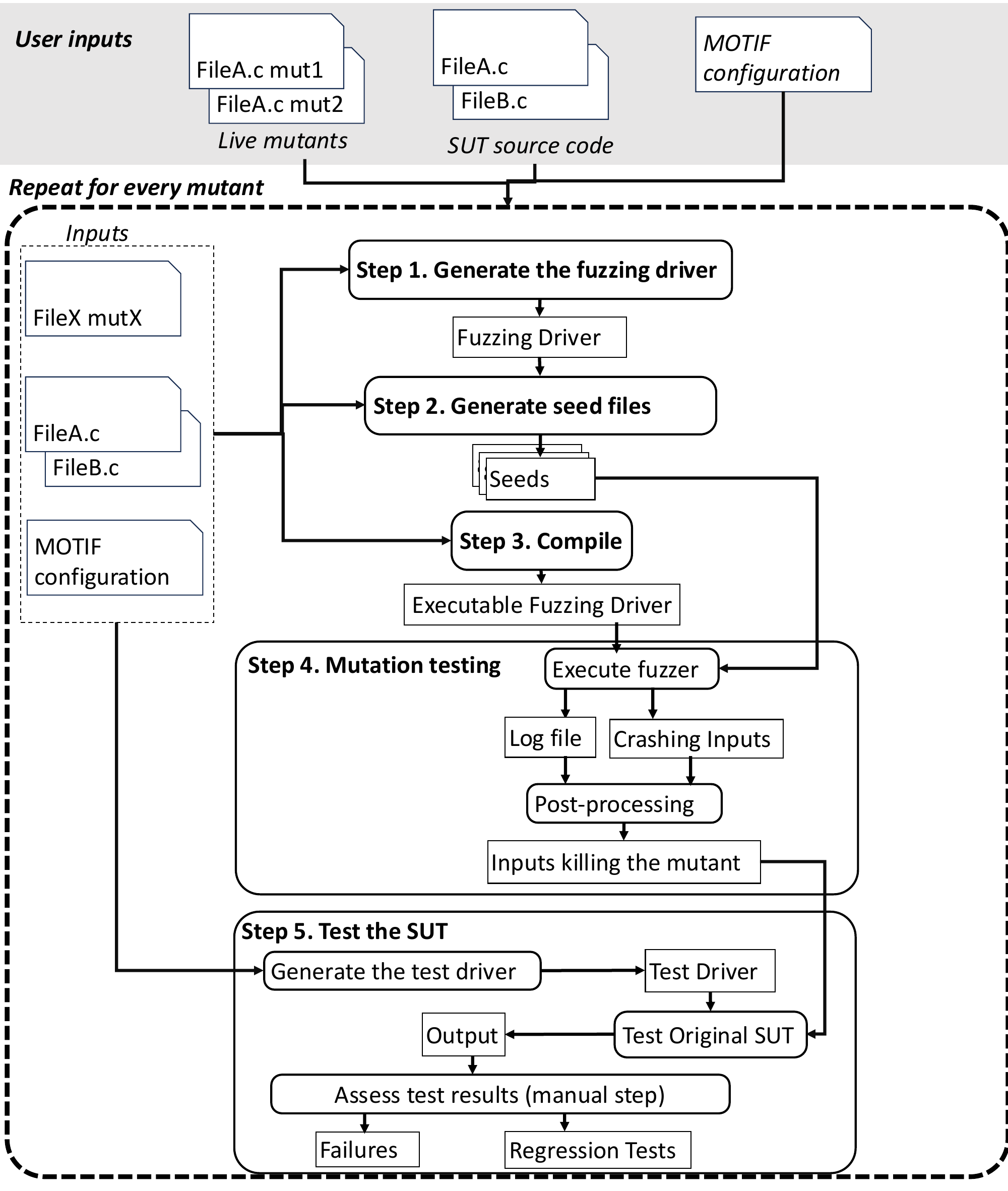}
\caption{The \MOTIF process.}
\label{fig:motif:process}
\end{center}
\end{figure}

Inspired by the work of Holling et al.~\cite{holling2016nequivack}, \MOTIF aims to identify a set of test inputs that lead to different outputs when given to both the original and mutated functions. To achieve such objective with fuzzing,  \APPR generates a fuzzing driver for each mutated function. The fuzzing driver reads the input data generated by the fuzzer and supplies it as arguments to both the original and mutated functions.  Finally, the fuzzing driver compares the outputs from both functions. If the outputs differ, the mutant is considered killed. 

Our intuition is that fuzzers might be 
effective at killing mutants because, by invoking both the original and the mutated functions within the same fuzzing driver, we can leverage the bucket-based fuzzing strategy to cover the different branches in the two functions in  diverse ways, thus reaching those program states that enable killing  mutants.
Essentially, the bucket-based fuzzing strategy may help kill mutants by preserving, during test generation, inputs that lead to incorrect intermediate states but do not kill the mutant (i.e., they do not meet the sufficiency condition). 
Subsequent iterations of the fuzzer’s evolutionary search process (see Section~\ref{sec:background:fuzzing}) may modify these inputs so that they not only reach an incorrect intermediate state but also meet the sufficiency condition.
Indeed, if differences in coverage between the original and mutated functions are observed, it indicates that the functions behave differently, resulting in different outputs and the mutant being killed~\cite{schuler2013covering,schuler2009efficient,Oscar:TSE}.
Additionally, significant differences in coverage lead to new buckets being covered, and since fuzzing favors inputs that cover new buckets, it indirectly leads to inputs that kill mutants. 
We leave to future work the extension of fuzzers with dedicated strategies; for example, instead of measuring the coverage of the mutated function, the fuzzer could measure the difference in coverage between  the original and the mutated function, and use this information to prioritize the inputs in the fuzzer queue (e.g., testing inputs that leads to larger differences first).

\APPR creates all the necessary scaffolding to test both the original and mutated functions, and to compare their outputs. Specifically, \APPR follows the workflow illustrated in Figure~\ref{fig:motif:process}, which comprises the five steps detailed below.

\APPR takes as input a set of mutants (source files) to be killed; each mutant matches the original source file except for the statements modified by a mutation operator. 
The \APPR Steps in Figure~\ref{fig:motif:process} are repeated for each mutant. 
However, Steps 1 and 2 can be executed only once for all the mutants belonging to the same function; indeed, the structure of the input and output data of a function is not changed by mutation---we do not target interface mutation~\cite{delamaro2001interface}.

\subsection{Step 1 -- Generate the fuzzing driver}
\label{sec:motif:step1}
In Step 1, MOTIF relies on the \emph{clang} static analysis library~\cite{CLANG} to analyse the SUT and determine the types of the parameters required by the function under test. This information is then used to generate a fuzzing driver for mutation testing; an example fuzzing driver for the function \texttt{T\_POS\_IsConstraintValid} belonging to our \ASNLib{} case study subject is shown in Listing~\ref{asn_driver}. The fuzzing driver renames the mutated function by adding the prefix \emph{mut\_}.

\input{listings/driver.tex}

The fuzzing driver contains two sets of variables (Lines 5-7 and 8-10) whose types match the parameters of the function under test and are provided as input to both 
the original and the mutated function. In our example, it declares a \texttt{struct T\_POS} and an \texttt{int} variable. 
These two sets of variables are then assigned by performing a byte-by-byte copy of a same portion of  input file provided by the fuzzer (Lines 16-17 and 23-24\RESUB{R1.a}{, achieved by the function \texttt{get\_value}}); \APPR ensures to copy a number of bytes to match the size of the assigned variable. If the input file provided by the fuzzer is shorter than required (the file modifications performed by fuzzers include shortening files), \APPR extends it with random data (Line~2). Additionally, the fuzzing driver declares the variables required to store the functions' return values (Lines 11 to 13).

The original and the mutated functions are then invoked (Lines 19 and 26).
The fuzzing driver then compares the outputs generated by both functions (Lines 28-31). Unfortunately, in C and C++, distinguishing between input and output parameters is complicated by the presence of pointer and reference arguments. Additionally, determining input parameters through data-flow analysis is impractical, as it requires preprocessing the SUT with a static analysis framework (e.g., LLVM~\cite{LLVM}), which is often not feasible for CPS software~\cite{Oscar:TSE}. 
Therefore, we use a straightforward approach to compare outputs which consists of comparing all parameters and return values of the original and mutated functions; such approach does not lead to incorrect mutant killing because input parameters remain unmodified. 
For pointers, we compare the data they point to (e.g., an \texttt{int} instance for \texttt{int*}). If the pointer is used as an array, the end-user can specify the expected length of the array, so the array data can be compared.  When arrays are inputs to the function under test, the end-user may not need to provide the length, as \APPR automatically generates arrays with a default length of 100. If the function under test dynamically allocates arrays, the end-user should specify the minimal possible length (e.g., an array of length one) to avoid false positives from out-of-bounds readings. For data structures with pointer fields, the pointed data length and initialization procedure can be specified. If the outputs differ, the fuzzing driver halts execution with an abort signal (Line 35 in Listing 1), allowing the fuzzer to detect the aborted execution and store the input file. MOTIF then stops the fuzzer because the mutant has been killed.

\input{tables/seeds.tex}

\subsection{Step 2 -- Generate seed files}

In Step 2, \APPR creates seed files based on the input parameter types for the function under test. These seed files are used by the fuzzer to initiate the testing process. Typically, fuzzers are executed with seed files that correspond to typical inputs for the SUT. In our approach, we automatically generate seed files that contain enough bytes to populate all input parameters with values that cover basic cases. Specifically, for each primitive type, we have identified three representative seed values for typical input partitions, as shown in Table~\ref{tab:seedValues}. For instance, for numeric values, we provide zero, a negative number, and a positive number. Using these seed values, \APPR generates up to three seed files for each fuzzing driver, ensuring that each seed value is covered at least once for every input parameter.

\begin{figure}[tb]
\begin{center}
\includegraphics[width=8.4cm]{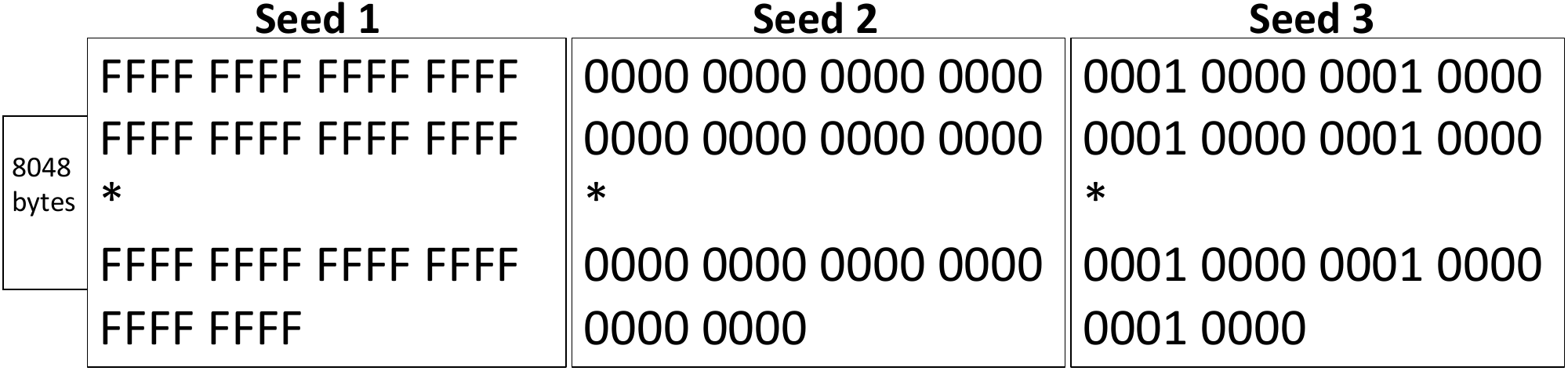}
\caption{Seed files generated for the fuzzing driver in Listing~\ref{asn_driver}.}
\label{fig:seedFiles}
\end{center}
\end{figure}

Example seed files for function \texttt{T\_POS\-\_IsConstr\-aint\-Valid} are provided in Figure~\ref{fig:seedFiles} (type definitions in Listing~\ref{asn_types}).
\MOTIF can also generate seed files for complex input types. For instance, 
the \texttt{struct T\_POS} received as input by 
function \texttt{T\_POS\_IsConstraintValid} consists of an \texttt{enum} (named \emph{kind}), which specifies the type of data stored inside the rest of the struct, and a union (named \emph{u}), which is sufficiently large to contain the data for all the data types selectable with the variable \emph{kind}. \MOTIF treats such struct as an \texttt{int} array thus filling it with the seeds \emph{0xFFFFFFFF}, \emph{0x00000000}, and \emph{0x00000001}. 
The first four bytes in the seed files (see Figure~\ref{fig:seedFiles}) belong to the \texttt{enum} item \emph{kind}, and are filled with the seed values of the \texttt{int} type. The same happens for the \texttt{union} field \emph{u} but, since the \texttt{union} has a size of 8,052 bytes (size of \emph{subTypeArray} with 4 bytes padding\footnote{https://research.nccgroup.com/2019/10/30/padding-the-struct-how-a-compiler-optimization-can-disclose-stack-memory/}), \MOTIF repeats the same set of four bytes 2,013 times. The last four bytes belong to the second parameter of \texttt{T\_POS\_IsConstraintValid}, the \texttt{int} \emph{*\_pErrCode}.

\input{listings/types.tex}

\subsection{Step 3 -- Compile the SUT}

In Step 3, \APPR produces an executable fuzzing driver by compiling the source code of the fuzzing driver, the mutated function, and the SUT using the fuzzer compiler, which is essential for gathering the code coverage information needed by the fuzzer. This step also compiles a false positive driver (see Section \ref{sec:stepFour} below).

\subsection{Step 4 -- Perform mutation testing}
\label{sec:stepFour}
In Step 4, \APPR runs the fuzzer to generate inputs for the executable fuzzing driver \REV{C2.3}{until either the time budget expires or a test input that can kill the mutant is found}\RESUB{R1.b}{; in our experiments we leverage the AFL++ fuzzer but the approach is generic and can work with any grey-box fuzzer for C/C++.} \CR{A.5}{The fuzzer keeps generating input files until it reports one or more crashes, after which MOTIF halts the fuzzer.} This process leads to the generation of fuzzing driver logs and crashing inputs (i.e., input files that caused a crash during the execution of the fuzzing driver). \CR{A.5}{Since fuzzers generate several input files from each input taken from the file pool, and since all of them are executed by the fuzzer, more than one crashing input may be reported.}

Fuzzing driver logs include checkpoints indicating the progress of testing (see Lines 18, 25, 28, 34, 37 in Listing~\ref{asn_driver}).
For each crashing input, \APPR processes the corresponding logs to distinguish between:
\begin{itemize}
\item Crashes occurring during the execution of the original function. They indicate either the presence of a fault in the original function or the violation of preconditions. We ignore these inputs because they do not correspond to inputs killing a mutant.
\item Crashes occurring during the execution of the mutated function. Since the mutated function is executed after the original one, we can safely conclude that the test inputs do not cause any crash in the original function. Therefore, the observed crashes indicate that the mutant introduced a fault that was exercised by the input. Thus, we can conclude such inputs kill the mutant.
\item Aborted executions due to the fuzzing driver determining that the mutant has been killed (see Line 35 in Listing~\ref{asn_driver}).
\end{itemize}

\MOTIF retains all test inputs that kill a mutant. 
However, the function under test may produce non-deterministic outputs meaning that, despite observed differences in outputs, the inputs may not have actually killed the mutant. 
 For instance, two consecutive invocations of a  function that reads and writes global variables may lead to different outputs even if the mutated statement is not exercised; consequently, the input suggested by the fuzzer would be a false positive.  
To minimize false positives, \MOTIF automatically re-executes every test input killing a mutant using a modified version of the fuzzing driver that
invokes the original function instead of the mutated function 
\REV{C2.8}{(see the post-processing of Step 4 in Figure 1)}.
If this false positive driver, as we refer to it, reports a difference in outputs of the two function calls, it implies that the function under test is non-deterministic and thus that the input may not kill the mutant. \MOTIF considers mutants exclusively killed by false positive inputs to be live. 
To kill mutants in functions that modify global state variables, the user must manually add, in the fuzzing driver between the two function calls, instructions to reset each state variable,  similar to other fuzzing approaches for unit and library testing (e.g., LibFuzzer~\cite{LibFuzzer}).

\subsection{Step 5 -- Test the SUT}

\input{listings/test}

In this Step, \APPR generates a test driver for the SUT. An example test driver for function \texttt{T\_POS\_IsCon\-straintValid} is shown in Listing~\ref{asn_test}. The test driver is identical to the fuzzing driver except that (1) it only invokes the original function (Line 10 in Listing~\ref{asn_test}), (2) instead of comparing outputs from two function invocations, it prints the output data generated by the original function (Lines 12 to 14), and (3) it includes assertions that compare the execution results of the original function with the expected results \REV{C2.4}{(Lines 15 to 18). The expected values used in the assertion statements are those observed during fuzzing for the original function and are stored in the test driver as a \texttt{char} array with a variable whose name starts with \texttt{ex\_} (Lines 2 to 4)}. 

The test driver is used to test the original SUT with the inputs that kill the mutant. \REV{C2.4}{To proceed with testing, engineers shall verify if the generated assertions are correct, which is a typical activity for automatically generated test cases. However, since our implementation of MOTIF stores expected values in binary form, we find that the easiest way to verify assertion correctness is by executing the test case and observing its outputs. 
Precisely, engineers shall observe the values printed out by the test driver for each variable of interest (Lines 12 to 14 in Listing~\ref{asn_test}); note that, sometimes, with complex data structures, the inspection of values is simplified through the use of a debugger. If the test outputs match the specifications, then the software behaves correctly; otherwise, the test case has discovered a fault. This scenario highlights one of the key advantages of mutation testing: When testing the SUT with inputs that detect simulated human mistakes (mutants), actual faults in the SUT are more likely to be discovered than with randomly selected inputs (e.g., in our experiments we discovered five faults in our subjects). When a fault is discovered, engineers need to fix the software but also update the assertions to reflect the expected behaviour instead of the behaviour observed during fuzzing.
Since the assertions generated by \APPR match the output observed during fuzzing, the test case is expected to pass even if the output does not match the specifications. If the test case does not pass, it means that the software behaves non-deterministically and MOTIF did not discover such nondeterminism automatically (it never happened in our experiments); in such case, engineers shall decide whether to keep the test case and update the assertions manually or simply discard the test case (e.g., because it does not kill the mutant). If the test case passes and the engineers confirm that the software outputs are correct, it can be integrated into the software test suite for future regression testing.}

In our test driver, the generation of print statements for structs and pointers is determined by the configuration of the fuzzing drivers. By default, all bytes within a struct are printed. When pointers are involved, if the end-user specifies the size of the data to which the pointers refer, the test driver prints the data pointed to, rather than the pointer value.

%% file: listings/driver.tex
\begin{figure}[h]
\begin{lstlisting}[style=CStyle, caption=Example fuzzing driver for the \ASNLib{} subject., label=asn_driver, mathescape=true]
int main(int argc, char** argv){
  load_file(argv[1]);  // load the input file and
  // extends the input with random data if needed

  /* Variables for the original function */
  T_POS origin_pVal;   // for the first parameter
  int origin_pErrCode; // for the second parameter
  /* Variables for the mutated function */
  T_POS mut_pVal;      // for the first parameter
  int mut_pErrCode;    // for the second parameter
  /* Variables for the return values */
  flag origin_return;  // for the original 
  flag mut_return;     // for the mutant

  /* Copy the input data to the variables for the original function */
  get_value(&origin_pVal, sizeof(origin_pVal), 0); 
  get_value(&origin_pErrCode,sizeof(origin_pErrCode),0); 
  log("Calling the original function");
  origin_return = T_POS_IsConstraintValid(&origin_pVal, &origin_pErrCode);

  /* Copy the same input data to the variables for the mutated function */
  seek_data_index(0); //reset the input data pointer
  get_value(&mut_pVal, sizeof(mut_pVal), 0); 
  get_value(&mut_pErrCode, sizeof(mut_pErrCode), 0); 
  log("Calling the mutated function");
  mut_return = mut_T_POS_IsConstraintValid(&mut_pVal, &mut_pErrCode);

  log("Comparing result values: ");
  ret += compare_value(&origin_pVal, &mut_pVal, sizeof(origin_pVal));
  ret += compare_value(&origin_pErrCode,&mut_pErrCode, sizeof(origin_pErrCode));
  ret += compare_value(&origin_return, &mut_return, sizeof(origin_return));

  if (ret != 0){
    log("Mutant killed");
    safe_abort();
  }
  log("Mutant alive");
  return 0;
}    
\end{lstlisting}
\end{figure}

%% file: tables/seeds.tex
\begin{table}[t]
\centering
\small
\caption{Seeds assigned to types}
\label{tab:seedValues}
\renewcommand{\arraystretch}{1.3}  
\begin{tabular}{
@{\hspace{1pt}}>{\raggedleft\arraybackslash}p{4em}@{\hspace{1pt}}
@{\hspace{1pt}}>{\raggedleft\arraybackslash}p{12em}@{\hspace{1pt}}
@{\hspace{1pt}}>{\raggedleft\arraybackslash}p{11em}@{\hspace{1pt}}
@{\hspace{1pt}}>{\raggedleft\arraybackslash}p{12em}@{\hspace{1pt}}}
\toprule
\multicolumn{1}{c}{\textbf{Type}} & 
\multicolumn{1}{c}{\textbf{Seed 1}} &
\multicolumn{1}{c}{\textbf{Seed 2}} &
\multicolumn{1}{c}{\textbf{Seed 3}} \\

\midrule
int& -1& 0& 1\\
Bool& False& True& \\
float& -3230283776.0& 0.0& 1072693248.0\\ double& 13826050856027422720.0& 0.0&
4602891378046628864.0\\
char& 0xFF& 0x00& 0x41\\
byte& 0xFF& 0x00& 0x41\\
ISO8601& 2145916800.999999999&
1970-01-01T00:00:00Z& 2038-01-01T00:00:00Z\\ 
\bottomrule
\end{tabular}%

\end{table}

%% file: listings/types.tex
\begin{figure}[tb]
\begin{lstlisting}[
style=CStyle, 
caption=Definition of \texttt{struct T\_POS}, 
label=asn_types, 
mathescape=true
]
typedef enum { T_POS_NONE,           longitude_PRESENT, 
               latitude_PRESENT,     height_PRESENT, 
               subTypeArray_PRESENT, label_PRESENT, 
               intArray_PRESENT,     myIntSet_PRESENT, 
               myIntSetOf_PRESENT,   anInt_PRESENT
} T_POS_selection;

typedef struct {
    T_POS_selection kind;
    union { asn1Real longitude; asn1Real latitude; 
            asn1Real height;    My2ndInt anInt; 
            T_POS_label label;  T_ARR intArray; 
            T_SET myIntSet;     T_SETOF myIntSetOf; 
            T_POS_subTypeArray subTypeArray; 
    } u;
} T_POS;
\end{lstlisting}
\end{figure}

%% file: listings/test.tex
\begin{figure}[tb]
\begin{lstlisting}[style=CStyle, caption=Example test driver for the \ASNLib{} subject., label=asn_test, mathescape=true]
// expected values
const unsigned char ex_pVal[8056] = {0x03,0x00,0x00,0x00,0x01,0x00,0x00,...};
const unsigned char ex_pErrCode[4] = {0x9B,0x01,0x00,0x00};
const unsigned char ex_return[1] = {0x00};

int main(int argc, char** argv){
  load_file(argv[1]); /* load the input file */
  // Declaration of variables and assignment with input file data missing to save space...
  /* Invoke the original function*/
  _return = T_POS_IsConstraintValid(&pVal, &pErrCode);
  /* Print output values of the original function */
  printf_struct("pVal (T_POS)=", &pVal, sizeof(pVal));
  printf("pErrCode (int) = %
  printf("return (flag) = %
  // assert statements  
  assert(0==compare_value((char *)&pVal, ex_pVal, sizeof(pVal)));
  assert(0==compare_value((char *)&pErrCode, ex_pErrCode, sizeof(pErrCode)));
  assert(0==compare_value((char *)&_return, ex_return, sizeof(_return)));
  printf("PASS\n");
  return 0;
}
\end{lstlisting}
\end{figure}

%% file: approach_cpp.tex
\section{MOTIF C++ extensions}
\label{sec:approach:cpp}

In this Section, we describe how we extended \APPR to test C++ programs. 
This extension is necessary to accommodate the four key object-oriented programming (OOP) properties -- abstraction, encapsulation, inheritance, and polymorphism -- and their impact on \RESUB{R2.1}{mutation testing with \MOTIF. Note that these problems are specific to C++ because mutation testing approaches for higher-level programming languages such as Java~\cite{MU2,fraser2011mutation} can leverage reflection to overcome them, as described below.}
For the sake of clarity, in this Section, we call \MOTIFC the version of \MOTIF that handles C code and was presented in Section~\ref{sec:approach}, and \MOTIFCPP the version of \MOTIF extended to deal with C++ features. Listing~\ref{cppExample} provides a running example.

\input{listings/cppExample}

\emph{Abstraction} is a key property of grouping variables and operations that are highly related. \emph{Classes} are the means to declare those variables (i.e., attributes) and operations (i.e., methods) as new data types. It reduces the complexity of developing software by letting users use the classes without knowing how its features are implemented. 
Instead, for testing, instantiating a class is necessary to exercise the class instance methods.
But \MOTIFC does not instantiate classes since, in C, operations are implemented by functions, which do not require any object to be instantiated.
In C++, static methods play the same role as C functions, however they are rare.
Consequently, \MOTIFC cannot be used to test most C++ methods.

\emph{Encapsulation} is realized by hiding (i.e., making \emph{private} or \emph{protected}) a subset of the instance variables and methods implemented by a class. Consequently, a subset of class methods (i.e., private and protected methods) cannot be exercised by \MOTIFC; indeed, functions that do not belong to a class (e.g., the main function of the \MOTIF fuzzing driver), cannot access the class private and protected members. 

\emph{Inheritance} enables a class to reuse (i.e., inherit) the methods declared by its superclass, if any. 
Inheritance complicates the unit testing automated by \MOTIFC because it leads to dependencies between superclasses and subclasses. Indeed, not only a subclass may execute superclass methods but the execution of a method declared by a superclass may trigger the execution of a method in a subclass. For example, the execution of method \texttt{ChargeableModule.isAutonomous()} in Listing~\ref{cppExample} triggers the execution of the superclass method \texttt{SatelliteModule.getAutonomy()}, which, in turn, triggers the execution of \texttt{ChargeableModule.getAvgConsumption()} and \texttt{ChargeableModule.getAvgCharging()}. As a consequence, the solution adopted by \MOTIFC to test mutants (i.e., creating a copy of the original method and renaming it) would prevent the testing of methods following such pattern; for example, renaming a mutant of method \texttt{getAvgConsumption()} in \texttt{ChargeableModule} as \texttt{mut\_getAvgConsumption()} would prevent its call from method \texttt{SatelliteModule.getAutonomy()}, thus rendering testing ineffective.

\input{tables/cpp}

\emph{Polymorphism} refers to multiple language features.
\emph{Ad hoc polymorphism} (i.e., method overloading) indicates that methods with a same name can process arguments of different types. Since these methods have distinct implementations, they do not affect \MOTIFC because it treats them as distinct functions. 
 \emph{Parametric polymorphism} (i.e., class templates) implies that certain methods can be exercised only if the object they belong to has been instantiated with appropriate parameters (i.e., a specific class name). This is the case for class \texttt{DynamicMission}, which requires the specification of the data type used to model mission objectives as in
 ``\texttt{mission = new DynamicMission<Landmark>();}''
    \emph{Subtype polymorphism} occurs when the code contains a call to a method belonging to a certain object, but the concrete method to be invoked is decided at runtime, depending on the object type. This is the case for method \texttt{ChargeableModule.isAutonomous()}, which invokes \texttt{Mission.duration()} and, at runtime, may execute the implementation of \texttt{duration()} provided by either \texttt{StaticMission} or \texttt{DynamicMission}. To test those methods, it is necessary to determine the type of the object to be  instantiated and passed as argument,  which is not supported by \MOTIFC.
    
To address the issues above, \MOTIFCPP implements a set of extensions that are summarized in Table~\ref{table:cppExtensions} and described below.

\input{listings/cppMutant}

To deal with \emph{inheritance}, \MOTIFCPP injects the mutated method into a copy of the original class. Examples are shown in Listing~\ref{cppMutant} and~\ref{cppMutantTwo}, which present two distinct mutants of class \texttt{ChargeableModule}. 
Note that in the case of Listing~\ref{cppMutantTwo}, the creation of a renamed class copy, instead of a renamed method copy as in \MOTIFC, ensures that the mutated method is executed when testing \emph{mut\_ChargeableModule.isAutonomous()}.  \RESUB{R2.1}{In Java, this issue is addressed in a similar but more efficient way~\cite{MU2}, by leveraging the classloader mechanism to load in memory both the hierarchy of the original class and its mutated version (one for each mutant), which is not feasible in C++ because of the lack of a classloader managed by a virtual machine.}

To deal with \emph{abstraction}, \MOTIFCPP creates a fuzz driver that, instead of invoking the mutated method, creates an instance of a test utility class to which it delegates the instantiation of the class under test and the execution of the method under test. An example fuzz driver is shown in Listing~\ref{cppDriver}; except for the instantiation of the utility class, it matches the fuzz driver created by \MOTIFC. A test utility class (i.e., \texttt{MOTIF::TestUtil}) is shown in Listing~\ref{cppUtility}. Test utility classes are automatically generated by \MOTIFCPP in Step 1, but often need to be manually modified as described in Section~\ref{sec:approach:cpp:manual}. 
A utility class includes state variables to hold the values to be passed as arguments to the constructor of the class under test and to the method under test (Lines 4-9 in Listing~\ref{cppUtility}).  
Still in \texttt{MOTIF::TestUtil}, two additional state variables are used to store references to the value returned by the method under test (Line 11) and the instance under test (Line 13). \RESUB{R2.1}{The solutions implemented by approaches targeting Java never need manual intervention because they can leverage reflection to determine what are the constructors and, recursively the constructors for their parameters, and just rely on automated trial and error to identify appropriate initializations~\cite{MU2,fraser2011mutation}.}

\input{listings/cppDriver}

\REVSTART
\REV{C2.4}{}
In \texttt{MOTIF::TestUtil}, methods \texttt{call} and \texttt{compare} take care of the execution of the method under test and the comparison of the results obtained by the original and the mutated method, respectively. Method \texttt{call} (Lines 15-27, Listing~\ref{cppUtility}), which executes the method under test, includes automatically generated instructions to populate parameters with fuzzed data (Lines 17 and 18). If the constructor of a function under test receives as input an abstract type, the instructions needed to construct the concrete type to be used shall be manually added (Lines 19-21). Last, method \texttt{call} invokes the method under test after instantiating the class under test and stores the returned value (Lines 23-26).
Method \texttt{compare} (Line 29-37) compares the results obtained by the original and the mutated class.
The fuzzing driver instantiates two utility classes for each original and mutated method (Lines 6 and 10, Listing~\ref{cppDriver}), and invokes the \texttt{call} method for them (Lines 7 and 11, Listing~\ref{cppDriver}). These results are compared by invoking the \texttt{compare} method (Line 13, Listing~\ref{cppDriver}).   
Its logic follows the one of the fuzzing driver generated by \MOTIFC.
\RESUB{R2.1}{Approaches for Java~\cite{MU2,fraser2011mutation} can leverage programming conventions and reflection to compare object states by leveraging the equals method, which is normally overridden to compare two instances of a class, or inspector methods, which can be identified through naming conventions or purity analysis (simpler to achieve on Java bytecode~\cite{Salcianu2005}).}
\REVEND

To deal with \emph{protected} and \emph{private} methods (i.e., \emph{encapsulation}), instead of modifying the class under test to specify the test utility class as a \emph{friend class}, we suggest the end-user to exercise, in the fuzzing driver, a public method that invokes the private/protected method under test. For example, to test mutant 2 in Listing~\ref{cppMutantTwo}, the end-user may still rely on the fuzzing driver shown in Listing~\ref{cppDriver}, because it tests method \texttt{isAutonomous}, which, in turn, executes method \texttt{getAvgConsumption} (i.e., the mutated method). \RESUB{R2.1}{In Java, protected and private methods can be invoked through reflection~\cite{Arcuri2017}.}

\subsection{Manual tuning of \MOTIF's test utility class}\label{sec:approach:cpp:manual}

Like \MOTIFC, \MOTIFCPP automatically identifies the variables to create in the test utility class based on method signatures. However, \MOTIFCPP requires the manual specification of the class to instantiate to deal with \emph{parametric} and \emph{subtype polymorphism}. For example, if a method declares parameters of an abstract type (e.g., \texttt{isAutonomous} receives one parameter of the abstract type \texttt{Mission}), the end-user needs to specify the additional parameters required to instantiate the concrete class to be used to exercise the method under test (Line 9, Listing~\ref{cppUtility}).  

Note that implementing such additional instructions is much less labor intensive than manually identifying the inputs that kill the mutant. Indeed, the proper initialization of the objects required for testing simply consists of the invocation of  constructor methods. In contrast, the identification of the inputs that kill a mutant require a comparison between the mutated and the original functions to determine which inputs makes their output differ. Further, we expect that end-users can easily update \MOTIF's test utility classes by copying the code used in unit test cases, except for assignments to primitive variables, which shall rely on the data provided by the fuzzer.
Indeed, in unit test cases, like in \MOTIF's test utility class, all the objects required by the method under test shall be instantiated; therefore, 
the instructions to be manually-specified in the test utility class can be a copy of the ones used in the manually implemented test cases, except for assignments to primitive variables. %

Last, we consider the need for the manual modification of test utility classes a limitation of the \MOTIF implementation rather than the approach itself.
Indeed, it is possible to integrate an additional lightweight static analysis step into \MOTIF to automate the generation of test drivers. Indeed, as demonstrated by approaches for the generation of fuzz drivers for API fuzzing~\cite{FuzzGen}, it is generally feasible  to implement a lightweight static analysis parser that automatically generates sequences of function calls by copying the content of existing programs (test cases, in our context); however,  implementing such a parser goes beyond the objective of this paper which, for the C++ part, is concerned with proposing and assessing a solution to determine when a mutated method is killed (i.e., through a test utility class and a copy of the mutated class) rather than delivering a production-ready tool. 

Concluding, although \MOTIFCPP delegates the handling of several C++ features to the manual tuning of the automatically generated test utility class, the automated generation of both the fuzzing driver and the test utility class still significantly decreases manual effort when compared to manual mutation testing.

%% file: listings/cppExample.tex
\begin{lstlisting}[style=CPPStyle, caption=C++ example classes. For simplicity\, we show only in-class method declarations and we hide some method declarations., label=cppExample, mathescape=true]
class SatelliteModule {
    private:
    /** Charge in watts per hour **/
    double getCharge(){ return power; }
    
    protected:    
    /** Average consumption in watts per hour **/
    double getAvgConsumption(){
        return consumedWatts()/elapsedTime();
    }
    
    public:
    /** Hours before being out of charge **/
    double getAutonomy(){
        return getCharge()/getAvgConsumption();
    }
};

class ChargeableModule : public SatelliteModule {
    private:
    /** Average watts charged per hour **/
    double getAvgCharging(){
        return chargedWatts()/elapsedTime();
    }

    protected:    
    double getAvgConsumption(){
        return super::getAvgConsumption()-getAvgCharging();
    }
    
    public:
    ChargeableModule(double consumptionBoundary, double chargeBoundary){
        this->consumptionBoundary = consumptionBoundary;
        this->chargeBoundary = chargeBoundary;
    }
    
    /** True if the mission lasts less than the autonomy of the module**/
    bool isAutonomous( Mission mission){
        return mission.duration() <= getAutonomy();
    }
};

/** Absract class for missions **/
class Mission {
    public:
    /** Duration in hours **/
    virtual double duration();
};

class StaticMission: Mission {
    private:
    double presetDuration;
    
    public:
    StaticMission(double presetDuration){
        this->presetDuration=presetDuration;
    }
    
    double duration(){
        return presetDuration-elapsedTime();
    }
};

template <class ObjectiveT> 
class DynamicMission: Mission {
    public:
    void objectiveAchieved(ObjectiveT T){
        //remove from the list of objectives to achieve
        this->objectives.remove(T);
    }
    
    double duration(){
        return objectivesToAchieve()/objectivesAchievedPerHour();
    }
};
\end{lstlisting}

%% file: tables/cpp.tex
\begin{table}[b]
    \centering
    \caption{\MOTIFCPP extensions to deal with C++ language features.}
    \label{table:cppExtensions}
    \renewcommand{\arraystretch}{1.5}  
    \begin{tabular}{m{12em}m{18em}}
    \toprule
       \multicolumn{1}{c}{\textbf{Language} \textbf{feature}} &
       \multicolumn{1}{c}{\textbf{\MOTIFCPP extension}}\\
    \midrule
        Inheritance & Copy mutant class.\\
        \hline
        Abstraction& Automatically generate test utility class.\\
        \hline
        Ad hoc polymorphism & Separate testing for each method.\\
        \hline
        Encapsulation           & \multirow{3}{*}{Manual editing of test utility class.}\\
        Parametric polymorphism& \\
        Subtype polymorphism& \\
    \bottomrule          
    \end{tabular}
\end{table}

%% file: listings/cppMutant.tex
\begin{figure}[b]
\begin{lstlisting}[style=CPPStyle, caption=Mutant class for a mutant of method \texttt{isAutonomous}., label=cppMutant, mathescape=true]
class mut_ChargeableModule : public SatelliteModule {

    //... same code as in ChargeableModule
    
    bool isAutonomous( Mission mission){
        //MUTANT 1: replaced <= with <
        return 
        mission.duration() < getAutonomy(); 
    }
};

\end{lstlisting}
\end{figure}

\begin{figure}[b]
\begin{lstlisting}[style=CPPStyle,  caption=Mutant class for a mutant of method \texttt{getAverageConsumption}., label=cppMutantTwo, mathescape=true]
class mut_ChargeableModule : public SatelliteModule {

    //... same code as in ChargeableModule

    protected:    
    double getAverageConsumption(){
        //MUTANT 2: removed '- getAverageCharging()'
        return super::getAverageConsumption();
    }

    //... same code as in ChargeableModule
};

\end{lstlisting}
\end{figure}

%% file: listings/cppDriver.tex
\begin{figure}[tb]
\begin{lstlisting}[style=CPPStyle, caption=Example \MOTIFCPP fuzz driver., label=cppDriver, mathescape=true]
int main(int argc, char** argv){
  load_file(argv[1]);  // load the input file and
  // extends the input with random data if needed

  log("Calling the original function")
  auto origin = new MOTIF::TestUtil<ChargeableModule>();
  origin->call();

  log("Calling the mutated function")
  auto mutant = new MOTIF::TestUtil<mut_ChargeableModule>();
  mutant->call();

  if (orig->compare(mut) != 0){
    log("Mutant killed");
    safe_abort();
  }
  log("Mutant alive")
}  
\end{lstlisting}
\end{figure}

\begin{figure}[tb]
\begin{lstlisting}[style=CPPStyle, caption=Example utility class., label=cppUtility, mathescape=true]
namespace MOTIF{
template<typename TypeA>
class TestUtil{
  //parameters for the constructor
  double ctor_1;
  double ctor_2;
  //parameters for the method under test
  Mission param_1;
  int manual_1;
  //return value
  bool _return;
  //instance
  TypeA* object;
  
  void call(){
    //populate variables with fuzz data
    get_value(&ctor_1, sizeof(ctor_1) );
    get_value(&ctor_2, sizeof(ctor_2) );
    //instructions for parameters of an abstract type
    get_value(&manual_1,sizeof(manual_1));//added
    param_1 = new StaticMission(manual_1);
    
    //instantiate the class under test
    object = new TypeA(ctor_1, ctor_2);
    //invoke the method under test
    _return = object->isAutonomous(param_1);
  }
  
  template<typename TypeB>
  int compare(TestUtil<TypeB> *rhs){
    int ret = 0;
    ret += compare_value(&this->ctor_1,  &rhs->ctor_1);
    ret += compare_value(&this->ctor_1,  &rhs->ctor_2);
    ret += compare_value(&this->param_1, &rhs->param_1);
    ret += compare_value(&this->_return, &rhs->_return);
    return ret;
  }
};  
}
\end{lstlisting}
\end{figure}

%% file: evaluation-new.tex
\section{Empirical Evaluation}
\label{sec:evaluation}

We address the following research questions:

\emph{RQ1. What fuzzer configuration leads to best results in MOTIF?}
Fuzzing results can vary depending on the configurations applied to the fuzzers, which include the selection of compilers, sanitizers, and coverage metrics.
Executables' speed depend on compilers' code optimization capability and affect fuzzing effectiveness; indeed, quicker executions lead to more inputs being tested and, likely, to more mutants being killed.
Sanitizers are helpful in detecting invalid behaviors, such as memory overflow and arithmetic overflow during software execution; consequently, they may facilitate mutants detection (e.g., mutants leading to such invalid behaviours) although their overhead may reduce the number of generated test inputs and, in turn, mutation testing effectiveness. 
Coverage metrics are used to determine when an input should be kept in the fuzzing queue, and
can affect input generation and mutants killing; for example, specific metrics may facilitate the discovery of unexplored paths, which may include the mutated instructions. 
Through this research question, we 
aim to determine what configurations would be optimal for \MOTIF.

\emph{RQ2. How does mutation testing based on fuzzing compare to mutation testing based on symbolic execution, for software where the latter is applicable?}
Since certain CPS units may still satisfy the assumptions of SE approaches (e.g., absence of floating-point instructions and black-box components), we aim to assess what approach performs better in such cases.

\emph{RQ3. How does \MOTIF's seeding strategy contribute to its results?}
\MOTIF kills mutants either through the generated seeds or through fuzzed inputs; we therefore aim to assess how the two strategies individually contribute to \MOTIF results in order to determine if fuzzing is indeed useful.

\emph{RQ4. Does hybrid-fuzzing improve the effectiveness of MOTIF?}
We aim to assess hybrid-fuzzing approaches because they demonstrated to be effective in increasing the code coverage obtained through fuzzing and thus may improve mutants detection.  Fuzzing can be an advantage to exercise conditions that are easy to satisfy (e.g., branch conditions controlled by a relational operator in a basic clause, such as \texttt{x>0}) whereas symbolic execution helps satisfy narrow branch conditions  (e.g., joining multiple clauses, such as \texttt{x>0 \&\& x<3})~\cite{yun2018qsym}.

\emph{RQ5. Is it cost-effective to reuse mutant-killing inputs?}
Since mutants are often redundant \footnote{
A mutant is considered redundant if it can be killed by the same input that kills another mutant, even though they may differ syntactically~\cite{shin2017theoretical}.}, multiple mutants can be killed by the same inputs. Consequently, inputs generated by MOTIF that successfully kill some mutated versions of a function might be used to test other mutated versions that remained live. However, the execution of additional inputs increases mutation testing time. This research question aims to assess the tradeoff between mutation score improvement and increased testing time due to reusing mutant-killing inputs.

\subsection{Subjects of the study}
\label{sec:evaluation:casestudy}
\input{tables/caseStudies}

To address our research questions, we considered software deployed on space CPS (satellites) currently in orbit.
This included (a) \MLFS{}, the Mathematical Library for Flight Software~\cite{MLFS}, which complies with the ECSS criticality category B~\cite{ecss40C,ecss80C}, (b) \UTIL{}, which is a utility library developed by GomSpace and used in NanoSatellites, 
(c) \ASNLib{}, a serialization/deserialization library,
(d) \SAIL, a subset of the control software of a micro-satellite developed by LuxSpace to track ships worldwide,
and (e) Sentinel-5 UVNS L1b Prototype Processor (\Sentinel, for brevity), which is ground software developed by Huld to process radiation data received from the Sentinel 5 satellite instruments.
\ASNLib{} has been generated with ASN1SCC from a test grammar provided by ESA. ASN1SCC is a compiler that generates C/C++ code suitable for low resource environments~\cite{ASN1CC,ASN1:paper}.
All the subjects are implemented in C except \Sentinel, which is implemented in C++.

Our software subjects are provided with test suites whose code coverage is reported in Table~\ref{table:caseStudies}.
Most test suites do not achieve 100\% statement coverage because they include components that are tested with specific hardware not available to us; therefore, we generated mutants only for the covered statements.
We generated mutants with MASS~\cite{Oscar:TSE,MASSTOOL}; specifically, we rely on all the mutation operators supported by MASS \RESUB{R1.d}{(i.e., the sufficient set~\cite{offutt1996experimental} and the deletion set~\cite{delamaro2014designing,andrews2005mutation})}, which proved effective in previous experiments on similar subjects. \REV{C1.4,3.2}{The mutation operators implemented by MASS are shown in Table~\ref{table:operators}. Though MASS does not include mutation operators specific to object-oriented programming languages~\cite{delgado2017assessment}, it is the only tool that enabled scalable mutation analysis in all our subjects, based on our previous experiments~\cite{Oscar:TSE}; further, including C++-specific operators would have led to a different setup across subjects.} 
\REV{C1.4, C3.1}{In addition, MASS automatically excludes} mutants that are identified as equivalent or duplicate according to trivial compiler equivalence methods\footnote{\REV{C3.1}{The percentage of equivalent and duplicate mutants excluded by MASS before computing the MS is significant: 23.93\% for \MLFS{}, 28.00\% for \UTIL{}, 26.19\% for \ASNLib{}, 36.10\% for \SAIL{}, 23.29\% for \Sentinel.}}~\cite{Oscar:TSE}. The last column %
in Table~\ref{table:caseStudies} provides the mutation score (MS) for our case study subjects; it corresponds to the proportion of mutants detected by the test suite. The highest mutation score is observed with MLFS, whose test suite achieves MC/DC adequacy~\cite{chilenski1994applicability}. The lowest mutation score is observed with \ASNLib{}, which is automatically generated by ASN1SCC using a grammar-based approach~\cite{ASN1:paper}. Our subjects' mutation scores are in line with empirical investigations reporting mutation scores ranging from 55\% to  95\%~\cite{Ramler2017,delgado2018evaluation}, for CPS software.

\input{tables/massMutants}

To perform test data generation, we rely on the mutants not killed by the original test suites. We assume that the live mutants are not equivalent (i.e., produce the same outputs for every input) to the original software. Although this could be an under-approximation, it does not introduce bias in the comparison between the different approaches considered (i.e., \SEMUp and \MOTIF configured with different fuzzers and options) because all of them cannot kill equivalent mutants. Further, two mutants $m_a$ and $m_b$ can also be duplicates (i.e., they lead to the same outputs for every input) or subsumed (i.e., $m_a$ is killed by a superset of the test cases killing $m_b$). 
However, the identification of test inputs that kill mutants is a precondition to determine if mutants are duplicate or subsumed~\cite{Shin:TSE:DCriterion:2018}; for this reason, including duplicate and subsumed mutants should not introduce bias in the comparison of the approaches. In other words, a mutation testing approach should easily kill mutants that are either duplicates or subsume other killed mutants; if it does not happen, it is correct to penalize such an approach in the empirical evaluation.

\input{tables/distMutants}

\REV{C2.10,C3.2}{Table~\ref{table:distMutants} reports the number of live mutants for each subject and the distribution of mutants across different mutation operators, as generated by MASS. The total number of mutants per subject is 437 for \UTIL{}, 1,347 for \ASNLib{}, 3,891 for \MLFS{}, 581 for \SAIL{}, and 99 for \Sentinel{}. While these mutants show that all types of mutation operators are covered, certain operators, such as ICR and UOI, contribute a larger number of mutants than others. This imbalance arises from both the nature of the mutation operators and the characteristics of the subjects. In particular, when test suites inadequately cover boundary conditions, mutants generated through small increments or decrements (e.g., by adding or subtracting 1) are more likely to survive. For example, in our case, many mutants deal with arrays. Modifying array indices by +1 or -1 often does not significantly affect program output, allowing such mutants not to be detected by the test suite.}

For \UTIL{}, \MLFS{}, and \SAIL{}, we configured \MOTIF to generate arrays of a specific size %
 and, for void pointers (i.e., \texttt{void *}), to create variables or arrays considering the expected data type.
Regarding \Sentinel, the relatively limited number of mutants considered is due to the need for the manual editing of most of the fuzz drivers generated by \MOTIF (see Section~\ref{sec:approach:cpp}). %
Indeed, advanced C++ features, including abstract classes, templates, friend functions, and unnamed namespaces, as well as the standard library implementations, render the drivers automatically generated by \MOTIF ineffective. For example, \texttt{vector<>}, one of the standard template libraries (STL), uses pointers  to trace the stored  data, which renders the memcopy-based solution adopted by \MOTIF to instantiate objects inappropriate (pointer addresses cannot be filled with data generated by the fuzzer). 
To address such limitation, we manually modified the fuzzing drivers automatically generated by \MOTIF. Precisely, in total, to kill 99 mutants, \MOTIF generated 25 drivers, 21 of which had to be manually modified. However, 14 drivers required a simple modification such as instantiating a template class (e.g., \texttt{vector<int>}), while seven drivers required an initialization step for the class under test, which we copied from the unit test cases provided by S5 developers. \RESUB{R1.e}{In addition, recall that, as discussed in Section~\ref{sec:approach:cpp:manual}, the need for such manual modifications is a limitation of the \MOTIF implementation, not the approach, and can be overcome through an improved static analysis step. Further, compared to manual mutation testing, the automated generation of both the fuzzing driver and the test utility class still significantly decreases manual effort.}

\subsection{Experimental setup}
\label{sec:evaluation:setup}

We performed our experiments using a prototype implementation of \MOTIF~\cite{MOTIF,MOTIFTool}.

\emph{To address all our RQs}, as fuzzer for \MOTIF, we selected \texttt{AFL++} because it is the fuzzer that performed better in terms of code coverage, according to recent benchmarks in the literature~\cite{FuzzBenchPaper,FuzzbenchReport,GreenFuzz}; moreover, along with HonggFuzz~\cite{honggfuzz}, it is the fuzzer that maximizes fault coverage in another recent benchmark~\cite{ICST:22:Sarro}.

\emph{To address RQ2}, and compare \MOTIF with a mutation testing approach relying on symbolic execution, we modified the \MOTIF pipeline to enable test generation with \SEMU; we call such pipeline \SEMUp. At a high level, \SEMUp follows the same steps of \MOTIF, with differences concerning how input and output variables are 
defined.

For \SEMUp, in Step 1, we generate \SEMU drivers enabling symbolic execution. An example \SEMU driver generated for function \texttt{T\_POS\_IsCon\-straint\-Valid} is shown in Listing~\ref{semu_driver}. These drivers must specify what are the input parameters to be treated symbolically (see Line 8 in Listing~\ref{semu_driver}), a task performed by the end-user. \SEMU drivers do not include explicit comparisons between the outputs of the mutated and the original function because such comparison is handled by \SEMU when symbolically executing the original and the mutated functions in parallel (see Section~\ref{sec:back:cs}). Precisely, the \SEMU driver invokes only the function under test and prints to standard output the data values that should be considered to determine if a mutant has been killed. Similar to \MOTIF, \SEMU also requires end-users to manually specify how to process data values belonging to data structures referenced with pointers.  
\SEMUp does not include a Step for the generation of seed input (i.e., \MOTIF's Step 2). It includes a step (corresponding to \MOTIF's Step 3) to compile the mutated function and the \SEMU drivers with LLVM, followed by a step, corresponding to \MOTIF's Step 4, for the execution of \SEMU and the processing of its logs to determine killed mutants.

\input{listings/semuDriver}

\emph{To address RQ3}, to perform hybrid fuzzing, we selected \texttt{SymCC}~\cite{SymCC}, whose performance is similar to that of the more recent tools Fuzzolic, SymSAN, and SymQEMU~\cite{SYMSAN}.
We excluded SymSAN and its recent extension Marco~\cite{Marco:ICSE:2024} because SymSAN requires 64-bit compilation, which is not feasible with some of our subjects. We excluded Fuzzolic and SymQEMU since they rely on a version of QEMU that cannot run on some of our subjects (e.g., \SAIL). 
Therefore, in our response to RQ3, the benefits of hybrid fuzzing can be considered to be a lower bound. 

To leverage hybrid fuzzing, we extended \MOTIF's Step 3 to also compile the fuzzing driver and SUT with \texttt{SymCC}'s compiler, in addition to \texttt{AFL++}'s. In Step 4, \MOTIF executes the two compiled executable drivers together, following 
\texttt{AFL++} procedures for hybrid fuzzing with \texttt{SymCC}\footnote{Guideline for SymCC: 
\hyperlink{https://github.com/AFLplusplus/AFLplusplus/tree/stable/custom\_mutators/symcc}{https://github.com/AFLplusplus/AFLplusplus/tree/ stable/custom\_mutators/symcc}}.
When \texttt{AFL++} discovers new interesting inputs according to its coverage metrics, \texttt{SymCC} is triggered to perform symbolic execution. The inputs found by \texttt{SymCC} are then merged into the queue of \texttt{AFL++}. %
Once either tool finds inputs killing the mutant, fuzzing stops and \MOTIF proceeds with Step 4's post-processing activity. %

In our experiments, we test mutants in parallel, by leveraging the multiple nodes of an HPC infrastructure~\cite{HPC}. This decision depends on the observation that leveraging parallelism (e.g., by relying on Cloud solutions) is 
a cost-effective choice in the case of complex CPS with many live mutants. \REV{C2.11}{Parallel testing of mutants is enabled by the fact that, in contrast to  MU2~\cite{MU2},} \MOTIF does not test live mutants with the same inputs. Therefore, \MOTIF can simply be deployed on different nodes and executed. 
However, we assess (RQ4) how MOTIF's effectiveness can be improved by reusing the mutant-killing inputs obtained in the different, parallel executions.

To account for randomness, we executed each approach (i.e., different \MOTIF setups and \SEMUp) ten times for each subject. For each mutant, we executed each approach for 10,000 seconds, which we determined, in a preliminary study, to be sufficient for \SEMUp to maximize the percentage of killed mutants. Precisely,  for fuzzing and hybrid fuzzing, we allocate 10,000 seconds of time budget and stop fuzzing once this is reached. But for \SEMUp, we allocate 10,000 seconds to the symbolic execution process, which means that, after the timeout, if the mutant has not been killed yet, \SEMUp still tries to generate test inputs using the path conditions traversed so far, which leads to an execution time for \SEMUp that is slightly higher than others (around 650 seconds more).

\REV{C1.3, C3.4}{In the following, we discuss the test effectiveness of each tool in terms of percentage of live mutants killed by the test cases generated by the tool. Further, we compare the tools' effectiveness in terms of difference in percentage (pp) of killed mutants:
$$\mathit{pp}\ \mathit{difference} = MS_{A} - MS_{B}$$
For example, with $MS_{A}=89.1$ and  $MS_{B}=87.2$, we report that $A$ outperforms $B$ by 1.9 pp.}

\input{evaluation/evaluation-RQ1}

\input{evaluation/evaluation-RQ2}

\input{evaluation/evaluation-RQ3}

\input{evaluation/evaluation-RQ4}

\input{evaluation/evaluation-RQ5}

\section{Threats to validity}

\REV{C3.3}{We discuss internal, conclusion, construct, and external validity according to standard practice~\cite{wohlin2024experimentation,feldt2010validity,Ralph2018}}.

\subsection{Internal validity.}%
To address threats to internal validity, 
we manually verified that \MOTIF, \SEMUp, and \MOTIF-\textit{Hybrid} correctly execute; and further, we manually inspected a large subset of the generated test cases and mutants killed in our experiments. Further, our false positive driver ensures that \MOTIF results are not affected by the presence of global variables or, more generally, non-determinism. Although we do not reset global state variables in fuzzing drivers, note that across all experiment runs, out of 32,875 mutants reported as killed by the fuzzing driver, only 125 were false positives (0.38\%), thus showing that non-determinism does not undermine the applicability of \MOTIF.

\REV{C3.3}{Additionally, a potential threat may arise due to our results depending on the specific fuzzer used in our experiments, AFL++. However,  AFL++}
is one of the best performing grey-box fuzzers according to recent benchmarks (see Section~\ref{sec:evaluation:setup}). Further, we assessed \MOTIF with a state-of-the art hybrid fuzzing solution. Although alternative hybrid fuzzers were not applicable to some of our subjects (see Section~\ref{sec:evaluation:setup}), they might have led to different and potentially better results; we leave their investigation to future work. 

\subsection{Conclusion validity.}

\REV{C3.3}{Threats to conclusion validity relate to statistical power of our results, invalid statistical test assumptions,  reliability of measurements, and random irrelevancies ~\cite{wohlin2024experimentation}.}

\REVSTART{}
To deal with non-determinism of the different tools considered in our experiments (\MOTIF, \SEMUp, \MOTIF-\textit{Hybrid}), we repeated the execution of each tool ten times. To discuss the significance of the differences across tools, we applied the non-parametric Mann Whitney U-test because the distribution of mutants killed is not known; U-test is also a common choice in software engineering papers~\cite{Arcuri:practicalGuide:ICSE:2015}.

Our measurements concern the number of killed mutants and execution time; for execution time we relied on the timing functions provided by the operating system, while for mutants killing we verified the correctness of our tools. We do not expect unreliable measurements. 

In our context, the only source of random noise might be the workload of the machines used to run the experiments, which may slow down the performance of some of the tools. To mitigate this threat, in addition to relying on high-performance computing nodes with guarantees for the provided service level, we manually inspected  the execution logs to ensure that no anomalies --- such as exceptions caused by the host environment --- were present.

\subsection{Construct validity.}
Construct validity concerns face, content, convergent, and predictive validity~\cite{Ralph2018}. The constructs considered in our work are effectiveness and cost. Effectiveness is measured through the percentage of live mutants killed, while cost is measured in terms of execution time. 

\emph{Face validity} concerns the selection of appropriate reflective indicators. For effectiveness, we rely on the percentage of live mutants killed, which is appropriate in this context because the final objective of the tools considered in our experiments is to maximize the percentage of killed mutants. For cost, we measured the time required to kill mutants; specifically, for RQ1 to RQ4 we set a fixed time budged for all the tools, for RQ5 we measured the time taken to re-execute all the inputs and the time taken to kill mutants. The time required to kill mutants is an appropriate measure of cost because it indicates how long engineers have to wait before improving their test suite and has an impact on the whole testing process. 

\emph{Content validity} concerns the adequacy of reflective indicators to cover the breadth of the construct. For effectiveness, we could consider the fault detection rate, measured as the proportion of real-world faults detected by the test cases generated by the different tools. However, fault detection rate can be significantly assessed only with a large set of real-world faults, which we do not have access to, for our subjects. 
For cost, we could consider the time required to configure the assessed tools, which we did not directly measure. To mitigate this threat, we report that RQ1 is not affected by such issue because all the tools have the same configuration time since they differ only regarding the selection of configuration parameters. In RQ2, we minimized the threat by constructing a pipeline for symbolic execution that resembles MOTIF's. RQ3 and RQ5 do concern the comparison of tools with different configurations. For RQ4, we report that hybrid fuzzing has a slightly more complex setup.

We are not affected by \emph{convergence validity} threats because we collected only one reflective indicator per construct. To address \emph{predictive validity}, we reported on significance of the differences when comparing different tools (RQ1, RQ2, RQ4).

\REVEND{}

\subsection{External validity.}
\REV{C3.3}{Threats to external validity may arise from the use of a limited set of subjects for our experiments, which may affect the generalizability of our results. To address this issue,}
we selected diverse software subjects that are installed and running on space CPS, including satellites currently in orbit: a mathematical library, a utility library, a data serialization component, onboard control software, and ground software. Since they implement a diverse set of features (mathematical operations, serialization, string and time utilities), they strengthen the generalizability of our results. Further, these types of software components are typical in many CPS systems including avionics, robotics, and automotive, thus suggesting the proposed approach may be useful in many sectors other than space.

\REV{C3.3,C1.4}{Another threat to external validity comes from the nature of the mutants considered in our experiments, which are constructed by MASS using sufficient and deletion operators. Note that these operators are the ones commonly implemented in other state-of-the-art mutation analysis tools~\cite{papadakis2019mutation}. Further, they were considered in experiments demonstrating correlation between mutation score and fault detection capabilities of test suites~\cite{papadakis2018mutation,Chekam:17}; consequently, test cases killing those mutants likely detect other mutants. We leave to future work the investigation of MOTIF effectiveness for other types of mutants, such as mutants generated using generative AI~\cite{muBert}, mutants focusing on cybersecurity properties of software~\cite{muVuln}, or mutants altering characteristics specific for object-oriented programs~\cite{delgado2017assessment}.}

%% file: tables/caseStudies.tex
\begin{table}[tb]
\caption{Subject artifacts.}
\label{table:caseStudies} 
\footnotesize
\vspace{-1.5em}
\begin{center}
\renewcommand{\arraystretch}{1.3}  
\begin{tabular}{lcrrrrr}%
\toprule 
      \parbox{2em}{\centering{\textbf{Subject}}}
    & \parbox{3em}{\centering{\textbf{Open-source}}}
    & \parbox{3em}{\centering{\textbf{LOC}}}
    & \parbox{4em}{\centering{\textbf{\# Test cases}}}
    & \parbox{5em}{\centering{\textbf{Statement coverage}}}
    & \parbox{6em}{\centering{\textbf{Mutation score (MS)}}}
    \\
\midrule 
    \UTIL{}& No& 10,576 & 201 & 83.20\% & 71.20\%  \\
    \ASNLib{}{}& Yes& 7,260 &  139 & 95.80\% &58.31\%  \\
    \MLFS{}{}& Yes& 5,402 &  4,042 & 100.00\% & 81.80\% \\
    \SAIL{} & No& 2,235 &  384 & 95.36\% & 65.36\%  \\
    \Sentinel{} & No& 54,696 &  36 & 62.23\% & 64.13\% \\

\bottomrule
\end{tabular}
\end{center}
\end{table}

%% file: tables/massMutants.tex
\newcommand{\op}{\mathit{op}}
\newcommand{\ArithmeticSet}{ \texttt{+}, \texttt{-}, \texttt{*}, \texttt{/}, \texttt{\%} }
\newcommand{\LogicalSet}{ \texttt{&&}, \texttt{||} }
\newcommand{\RelationalSet}{ \texttt{>}, \texttt{>=}, \texttt{<}, \texttt{<=}, \texttt{==}, \texttt{!=} }
\newcommand{\BitWiseSet}{ \texttt{\&}, \texttt{|}, \land }
\newcommand{\ShiftSet}{ \texttt{>>}, \texttt{<<} }

\begin{table}[tb]
\caption{Mutation operators implemented in MASS.}
\label{table:operators} 
\centering
\scriptsize
\begin{tabular}{|@{}p{2cm}@{\hspace{1pt}}|@{}p{9.7cm}@{}|}
\hline
\textbf{Operator} & \textbf{Description$^{*}$} \\
\hline
\hline
\multicolumn{2}{|c|}
{\emph{Sufficient Set}}\\
\hline
ABS               & $\{(v, -v)\}$	\\
\hline
AOR               & $\{(\op_1, op_2) \,|\, \op_1, \op_2 \in \{ \ArithmeticSet \} \land \op_1 \neq \op_2 \} $       \\
    			  & $\{(\op_1, \op_2) \,|\, \op_1, \op_2 \in \{\texttt{+=}, \texttt{-=}, \texttt{*=}, \texttt{/=}, \texttt{\%} \texttt{=}\} \land \op_1 \neq \op_2 \} $       \\
\hline
ICR               & $\{i, x) \,|\, x \in \{1, -1, 0, i + 1, i - 1, -i\}\}$           \\
\hline
LCR               & $\{(\op_1, \op_2) \,|\, \op_1, \op_2 \in \{ \texttt{\&\&}, || \} \land \op_1 \neq \op_2 \}$            \\
				  & $\{(\op_1, \op_2) \,|\, \op_1, \op_2 \in \{ \texttt{\&=}, \texttt{|=}, \texttt{\&=}\} \land \op_1 \neq \op_2 \}$            \\
			  & $\{(\op_1, \op_2) \,|\, \op_1, \op_2 \in \{ \texttt{\&}, \texttt{|}, \texttt{\&\&}\} \land \op_1 \neq \op_2 \}$            \\
\hline
ROR               & $\{(\op_1, \op_2) \,|\, \op_1, \op_2 \in \{ \RelationalSet \}\}$            \\
				  & $\{ (e, !(e)) \,|\, e \in \{\texttt{if(e)}, \texttt{while(e)}\} \}$ \\
\hline
SDL               & $\{(s, \texttt{remove}(s))\}$            \\
\hline
UOI               & $\{ (v, \texttt{--}v), (v, v\texttt{--}), (v, \texttt{++}v), (v, v\texttt{++}) \}$            \\   
\hline
\hline
\multicolumn{2}{|c|}{\emph{OODL}}\\
\hline
AOD               & $\{((t_1\,op\,t_2), t_1), ((t_1\,op\,t_2), t_2) \,|\, op \in \{ \ArithmeticSet \} \}$       \\ 
\hline
LOD               & $\{((t_1\,op\,t_2), t_1), ((t_1\,op\,t_2), t_2) \,|\, op \in \{ \texttt{\&\&}, || \} \}$       \\ 
\hline
ROD               & $\{((t_1\,op\,t_2), t_1), ((t_1\,op\,t_2), t_2) \,|\, op \in \{ \RelationalSet \} \}$       \\ 
\hline
BOD               & $\{((t_1\,op\,t_2), t_1), ((t_1\,op\,t_2), t_2) \,|\, op \in \{ \BitWiseSet \} \}$       \\ 
\hline
SOD               & $\{((t_1\,op\,t_2), t_1), ((t_1\,op\,t_2), t_2) \,|\, op \in \{ \ShiftSet \} \}$       \\ 
\hline
\hline
\multicolumn{2}{|c|}{\emph{Other}}\\
\hline
LVR			& $\{(l_1, l_2) \,|\, (l_1, l_2) \in \{(0,-1), (l_1,-l_1), (l_1, 0), (\mathit{true}, \mathit{false}), (\mathit{false}, \mathit{true})\}\}$           \\
\hline
\end{tabular}

$^{*}$Each pair in parenthesis shows how a program element is modified by the mutation operator on the left; we follow standard syntax~\cite{Kintis2018}. Program elements are literals ($l$), integer literals ($i$), boolean expressions ($e$), operators ($\op$), statements ($s$), variables ($v$), and terms ( $t_i$, which might be either variables or literals).
\end{table}

%% file: tables/distMutants.tex
\begin{table*}[tb]
    \centering
    \caption{The number of live mutants for each subject and their distribution for each mutation operator obtained from MASS.}
    \label{tab:results:rq1:sub}
    \label{table:distMutants} 
\resizebox{\textwidth}{!}{
    \begin{tabular}{c|rrrrr|r}
    
    \toprule
        \multicolumn{1}{c}{Operators} 
        & \multicolumn{1}{c}{\UTIL{}} 
        & \multicolumn{1}{c}{\ASNLib{}} 
        & \multicolumn{1}{c}{\MLFS{}} 
        & \multicolumn{1}{c}{\SAIL{}} 
        & \multicolumn{1}{c}{\Sentinel{}} 
        & \multicolumn{1}{c}{Total} \\
\midrule
	ABS	 &  6 (\phantom{0}1.37\%)	 &  20 (\phantom{0}1.48\%)	 &  232 (\phantom{0}5.96\%)	 &  0 (\phantom{0}0.00\%)	 &  5 (\phantom{0}5.05\%)	 &  263 (\phantom{0}4.14\%)  \\ 
	AOR	 &  7 (\phantom{0}1.60\%)	 &  0 (\phantom{0}0.00\%)	 &  369 (\phantom{0}9.48\%)	 &  201 (34.60\%)	 &  14 (14.14\%)	 &  591 (\phantom{0}9.30\%)  \\ 
	ICR	 &  56 (12.81\%)	 &  426 (31.63\%)	 &  809 (20.79\%)	 &  103 (17.73\%)	 &  39 (39.39\%)	 &  1433 (22.55\%)  \\ 
	LCR	 &  52 (11.90\%)	 &  18 (\phantom{0}1.34\%)	 &  96 (\phantom{0}2.47\%)	 &  5 (\phantom{0}0.86\%)	 &  0 (\phantom{0}0.00\%)	 &  171 (\phantom{0}2.69\%)  \\ 
	ROR	 &  115 (26.32\%)	 &  303 (22.49\%)	 &  348 (\phantom{0}8.94\%)	 &  48 (\phantom{0}8.26\%)	 &  7 (\phantom{0}7.07\%)	 &  821 (12.92\%)  \\ 
	SDL	 &  33 (\phantom{0}7.55\%)	 &  197 (14.63\%)	 &  196 (\phantom{0}5.04\%)	 &  117 (20.14\%)	 &  11 (11.11\%)	 &  554 (\phantom{0}8.72\%)  \\ 
	UOI	 &  108 (24.71\%)	 &  186 (13.81\%)	 &  1377 (35.39\%)	 &  83 (14.29\%)	 &  10 (10.10\%)	 &  1764 (27.76\%)  \\ 
\midrule
	AOD	 &  0 (\phantom{0}0.00\%)	 &  0 (\phantom{0}0.00\%)	 &  234 (\phantom{0}6.01\%)	 &  11 (\phantom{0}1.89\%)	 &  6 (\phantom{0}6.06\%)	 &  251 (\phantom{0}3.95\%)  \\ 
	LOD	 &  40 (\phantom{0}9.15\%)	 &  39 (\phantom{0}2.90\%)	 &  5 (\phantom{0}0.13\%)	 &  3 (\phantom{0}0.52\%)	 &  0 (\phantom{0}0.00\%)	 &  87 (\phantom{0}1.37\%)  \\ 
	ROD	 &  19 (\phantom{0}4.35\%)	 &  126 (\phantom{0}9.35\%)	 &  72 (\phantom{0}1.85\%)	 &  6 (\phantom{0}1.03\%)	 &  0 (\phantom{0}0.00\%)	 &  223 (\phantom{0}3.51\%)  \\ 
	BOD	 &  1 (\phantom{0}0.23\%)	 &  0 (\phantom{0}0.00\%)	 &  48 (\phantom{0}1.23\%)	 &  2 (\phantom{0}0.34\%)	 &  0 (\phantom{0}0.00\%)	 &  51 (\phantom{0}0.80\%)  \\ 
	SOD	 &  0 (\phantom{0}0.00\%)	 &  0 (\phantom{0}0.00\%)	 &  11 (\phantom{0}0.28\%)	 &  0 (\phantom{0}0.00\%)	 &  0 (\phantom{0}0.00\%)	 &  11 (\phantom{0}0.17\%)  \\ 
\midrule
	LVR	 &  0 (\phantom{0}0.00\%)	 &  32 (\phantom{0}2.38\%)	 &  94 (\phantom{0}2.42\%)	 &  2 (\phantom{0}0.34\%)	 &  7 (\phantom{0}7.07\%)	 &  135 (\phantom{0}2.12\%)  \\ 
\midrule
        Total	 
            &   437	 \hspace{3.9em}
            &  1347	 \hspace{3.9em}
            &  3891	 \hspace{3.9em}
            &  581	 \hspace{3.9em}
            &  	99	 \hspace{3.9em}
            & 6355   \hspace{3.9em}	\\
\bottomrule												
    \end{tabular}
}
\end{table*}

%% file: listings/semuDriver.tex
\begin{figure}[tb]
\begin{lstlisting}[style=CStyle, caption=Example \SEMU driver corresponding to the fuzzing driver in~Listing~\ref{asn_driver}., label=semu_driver, mathescape=true]
int main(int argc, char** argv){
    // Declare variable to hold function returned value
    _Bool result;
    // Declare arguments and make input ones symbolic
    T_POS pVal;
    int pErrCode;
    
    klee_make_symbolic(&pVal, sizeof(pVal), "pVal");
    // Call function under test
    result = T_POS_IsConstraintValid(&pVal, &pErrCode);
    // Print output data
    printf("pErrCode = %
    printf("result = %
    return (int)result;
}
    
\end{lstlisting}
\end{figure}

%% file: evaluation/evaluation-RQ1.tex
\subsection{RQ1 - Fuzzer configurations}

\subsubsection{Design}
Since we measure effectiveness in terms of percentage of killed mutants, to address RQ1, we compare the percentage of mutants killed when different fuzzer configurations are selected; specifically, we compare the configuration parameters belonging to three distinct dimensions:
compilers, sanitizers, and coverage metrics. For each configuration considered in our experiment, we execute MOTIF for at most 10,000 seconds for each mutant and track the time required to kill each mutant. We compare distinct approaches in terms of number of mutants killed (i.e., mutation score, \emph{MS}) over time and evaluate the significance of the difference by relying on the Mann–Whitney U-test, a non-parametric test.

The first comparison focuses on the impact of the compiler choice on \MOTIF's effectiveness. We consider two compilers: \texttt{GCC} and \texttt{Clang}. \texttt{GCC} is the default compiler for most software projects in several domains (e.g., space).
\texttt{Clang} is a recent compiler that is gaining popularity across industries and is particularly appealing for fuzzing because it integrates several optimization techniques leading to faster programs\footnote{AFL++ with LLVM: \hyperlink{https://github.com/AFLplusplus/AFLplusplus/blob/stable/instrumentation/README.llvm.md}{https://github.com/AFLplusplus/AFLplusplus/ blob/stable/instrumentation/README.llvm.md}}\footnote{Clang-Features and Goals: \hyperlink{https://clang.llvm.org/features.html}{https://clang.llvm.org/features.html}}.
We select \texttt{Clang} version 14, since it supports all sanitizers and coverage options in AFL++, and \texttt{GCC} version 11.

The second comparison is among sanitizers. We consider four sanitizers supported by AFL++: \texttt{ASAN}~\cite{Konstantin2012} which detects array out-of-bound and invalid memory access, \texttt{UBSAN}~\cite{UBSAN:Clang19}  which detects undefined behavior such as arithmetic overflow, \texttt{LSAN}~\cite{LSAN:Clang19} which focuses on heap memory leaks, and \texttt{MSAN}~\cite{Stepanov2015} which focuses on uninitialized memory reads. 
We do not consider the two other sanitizers supported by AFL++, Control Flow Integrity SANitizer (CFISAN) and Thread SANitizer (TSAN), because they focus on problems specific for object-oriented and concurrent programs that are unlikely caused by code mutations (indeed, the mutation operators considered for our experiments don't aim at introducing multi-threading faults nor breaking control flows by altering pointer to functions). 
To perform our experiment, we compile our subjects (fuzzing drivers and SUT) multiple times, one for each sanitizer, with \texttt{Clang} (the best compiler based on our results, as discussed below), and execute \MOTIF on each version. 
We compare the observed results (MS) with the ones obtained with a baseline consisting of the use of \texttt{Clang} without any sanitization options.

In the third comparison, 
we considered three \emph{coverage metric options} supported by AFL++\RESUB{R2.4}{, which is motivated by the fact that, since fuzzers are driven by code coverage, coverage metrics affect the fitness of inputs.
They are: (1) \texttt{LAF}~\cite{LAF:Blog}, which splits complex comparison operations into multiple conditions\footnote{LAF splits conditions with non-strict relational operators such as \texttt{if (a <= b){doA}} or \texttt{if (a >= b){doB}}) into compound conditions such as \texttt{if (a == b)\{ doA \} else if (a > b)\{ doA \}}). It also splits invocations of \texttt{strcmp}, to compare strings, into an appropriate set of nested if conditions. Last it splits \texttt{switch} blocks into nested if conditions performing comparisons by-by-byte.}, thus preserving inputs that exercise boundary cases differently than previous inputs; (2) \texttt{NGRAM}~\cite{Wang2019} which tracks the coverage of sequences of $N$ edges, thus preserving inputs that exercise different sub-paths in a program,} and context-sensitive coverage (\texttt{CTX})~\cite{Wang2019} 
which combines edge coverage with information on call points, 
thus enabling the fuzzer to distinguish between the execution of the same code from different calling contexts.
For \texttt{LAF}, we enabled the splitting of all the types of compound expressions supported by \texttt{LAF}, and for \texttt{NGRAM}, we set the parameter N to 2.
To compile our subjects, we rely on \texttt{Clang}. 

\REV{C2.6}{Last, to discuss complementarity across configurations, we compare the number of mutants uniquely killed or missed by each configuration.}

\input{evaluation/table_rq1}

\subsubsection{Results}
Table~\ref{tab:results:rq1:sub} reports \MOTIF's performance, for each subject, when different fuzzer configurations are selected. 
Each row shows the average percentage of live mutants killed by \MOTIF over ten runs after 10,000 seconds, for a given fuzzer configuration.
$p$-values are shown in brackets, capturing the statistical significance of the difference between the selected configuration and the \texttt{Clang} baseline; for the compiler selection assessment, $p$-values simply capture the significance of the difference between the results obtained using the two available options.
A green background highlights cases where the selected configuration significantly outperformed the baseline; a red background highlights the opposite. Cases not leading to any significant difference have a white background.

The compilers' comparison results show that
\texttt{Clang} outperforms \texttt{GCC} by 1.37 pp in \UTIL{}, 1.13 pp in \ASNLib{}, 2.91 pp in \MLFS{}, and 0.65 pp in \SAIL{}. \REV{C2.12}{For each of these four subjects, the results obtained with the two compilers are significantly different at every minute.}
The better performance of \texttt{Clang} is likely due to the fact that the number of inputs generated by AFL++ with \texttt{Clang} is 4 to 20 times higher than the one observed with \texttt{GCC}.
Indeed, search algorithms (i.e., AFL++) are more likely to explore a larger portion of the input space when they generate more inputs. The larger number of inputs generated with \texttt{Clang} can be explained by \texttt{Clang} generating executable programs (in our case, fuzz drivers) that are quicker to execute than \texttt{GCC} ones and therefore AFL++ can try more inputs with \texttt{Clang} than with \texttt{GCC}, for the same test budget. 
In contrast to the above, for \Sentinel{}, \texttt{Clang} performed worse than \texttt{GCC} by 0.30 pp but the difference is not statistically significant, enabling us to conclude that overall \textbf{\texttt{Clang} is the best choice for \MOTIF}.

The sanitizer results in Table~\ref{tab:results:rq1:sub}  
show that \MOTIF equipped with sanitizers tends to perform worse than the baseline, \MOTIF without any sanitizer. 
\texttt{ASAN} led to the highest number of cases with improvements; indeed, it slightly outperformed the baseline by 0.25 pp ($p < 0.05$) in \ASNLib{}  and by 0.33 pp ($p < 0.01$) in \SAIL{}. \texttt{UBSAN}, instead, led to the largest improvement, with a proportion of killed mutants higher than the baseline by 1.67 pp ($p < 0.01$) in \UTIL{}; however, it performs worse than the baseline for most of the subjects.

The primary reason why \MOTIF with sanitizers cannot outperform the baseline is the overhead of the instrumented code introduced by sanitizers to detect invalid behaviors. 
Such overhead reduces the number of inputs generated by AFL++, consequently decreasing the number of mutants killed, even though sanitizers may kill some mutants not killed without them.
In our investigation, the worst performance is obtained with \texttt{MSAN}, which led to 96.77\% fewer inputs than the baseline, on average, and showed significantly lower performance. The worst \texttt{MSAN} result is obtained with \Sentinel{}, likely because of poor support for \texttt{C++} libraries (e.g., \texttt{MSAN} led to crashes whenever it encounters a standard library function such as \texttt{std::map} and \texttt{std::set}). \texttt{ASAN}, instead, led to 61.5\% fewer inputs, on average, but enabled the detection of a large number of additional mutants thus compensating the loss due to execution cost; consequently, \texttt{ASAN} achieved results that are comparable to those obtained without sanitizer.
\texttt{ASAN} effectiveness in killing mutants mainly depends on its capability to detect out-of-bound memory accesses caused by mutants; such mutants are harder to detect otherwise because the \emph{sufficiency condition} (see Section~\ref{sec:background:mt}) is hard to meet without a sanitizer (the effect of the out-of-bound access shall affect the values assigned to an output variable). Further, \texttt{ASAN} prevents the generation of test inputs that kill mutants but are not valid; those are inputs that violate function preconditions and cause out-of-bound accesses in the original function under test, and are thus discarded by \MOTIF.
Different from \texttt{ASAN}, \texttt{UBSAN} prevents the testing of several mutants because it detects invalid behaviors in the original function (on average, 653.8 fuzzing drivers failed when executing the original function), which need to be fixed before proceeding with mutation testing.
A representative example is provided by \MLFS{}, where \texttt{UBSAN} detects several arithmetic overflows caused by shift operators applied to variables that may have negative values; \REV{C2.5}{MLFS developers reported that \texttt{UBSAN} findings are overall useful for software development (e.g., they considered them to avoid mistakes when refactoring the library for a newer version, called LibmCS~\cite{libMCS}) but they did not detect any fault (i.e., the overflows are not possible because of implicit function preconditions).}
Although \texttt{UBSAN} may have detected potential bugs in the SUT, by terminating the execution of the original function, \texttt{UBSAN} prevents mutation testing. \REV{C2.5}{For critical software, we therefore suggest using slightly modified versions of MOTIF fuzz drivers (e.g., invoking only the function under test), to identify possible bugs with \texttt{UBSAN}; however, assessing the effectiveness of this approach is out of the scope of this paper.} In summary, \textbf{\texttt{ASAN} is the only sanitizer that is beneficial to \MOTIF for two of the subjects} (i.e., prevents false positives without largely affecting effectiveness).

\emph{Coverage metrics results} in Table~\ref{tab:results:rq1:sub} show that their effectiveness  vary across subjects. 
Specifically, \MOTIF with \texttt{LAF} outperformed the baseline (i.e., \MOTIF with \texttt{AFL++} using the default coverage options) by 2.27 pp and 1.66 pp, with $p$-value $\leq0.01$, in \ASNLib{} and \MLFS{}, respectively. 
With these two subjects, \texttt{LAF} likely improves reachability (i.e., generating inputs that reach the mutated statement) because they present several conditional statements that can be split into simpler conditions; indeed, %
\ASNLib{} contains compound conditions that check input validity, while \MLFS{}, relies on several signed and floating point  comparisons. In \SAIL{}, \texttt{LAF} performed significantly worse than the baseline by 0.31 pp; however, the practical impact of such decrease in performance is very limited (2 fewer mutants killed, in the worst case).

\MOTIF with \texttt{NGRAM} and \texttt{CTX} outperformed the baseline in both \UTIL{} and \Sentinel{}. In \UTIL{}, we observe an improvement of 1.17 pp and 1.21 pp for \texttt{NGRAM} and \texttt{CTX}, respectively ($p$-value $\leq0.01$). In \Sentinel{}, the improvements are 0.91 pp and 1.01 pp, respectively ($p$-value $\leq0.01$). 
In \UTIL{}, the improvements are likely due to \UTIL{} including longer call sequences whose results depend on state variables. In \Sentinel{}, the improved performance
 is likely due to \texttt{C++} language features (i.e., overriding and polymorphism), resulting in method invocations that behave differently depending on the call context.
 In the worst case, \texttt{NGRAM} and \texttt{CTX}, decrease \MOTIF's performance by 0.30 pp (\texttt{NGRAM} on \ASNLib{}) and 
 0.14 pp (\texttt{CTX} on \MLFS), with an average of 4 to 6 fewer mutants being killed, a small impact in practice.
To summarize, \textbf{it is not possible to identify a coverage metric that works best with all the subjects; however, since the differences are limited, engineers may select the option that improves testing results in critical code units} (e.g., \texttt{LAF} because it improves MS in mathematical functions).

Although our results did not help identifying a sanitizer and coverage metric configuration that provides effectiveness improvements across all subjects, 
for the sake of simplifying the remaining experiments, we identify a configuration for \MOTIF that is likely to provide improvements in most subjects. It consists of combining \texttt{Clang}, \texttt{ASAN}, and \texttt{LAF} and we call such configuration \texttt{Best}. 
We selected \texttt{ASAN} because it performs similarly to the baseline in terms of percentage of killed mutants but also prevents the generation of inputs that are invalid and cause memory access errors. We selected \texttt{LAF} because it leads to the largest number of killed mutants, across all subjects (3283 versus 3186 for NGRAM and 3190 for CTX), and further leads to the largest effectiveness improvement (+2.27 pp), and also improves the subject with the most effective test suite (i.e., \MLFS, which achieves MC/DC and 81.80\% MS).

\input{evaluation/fig_rq1}

To ensure that the selected configuration options do not interfere, we performed ten additional executions of \MOTIF configured with the \texttt{Best} settings, with each subject, and compare the observed results with \MOTIF relying on the default \texttt{AFL++} options (i.e., relying on \texttt{GCC} without additional sanitizers or coverage metrics). Figure~\ref{fig:results:rq1} shows the percentage of mutants killed by \MOTIF after each second, when using \texttt{GCC} and \texttt{Best}, for each subject. 
Each line captures the average percentage observed in ten runs, with the shaded area capturing the upper and lower bounds across those runs.
The vertical dashed line shows the time budget for the experiment.
With \texttt{Best}, the percentages of killed mutants are 49.84\% for \UTIL{}, 88.74\% for \ASNLib{}, 39.85\% for \MLFS{}, 39.50\% for \SAIL{}, and 82.53\% for \Sentinel{}.
These performance results are significantly better than those of our baseline (i.e., \MOTIF with \texttt{GCC}) for four subjects: it increases the percentage of killed mutants by 1.53 pp in \UTIL{}, 3.30 pp in \ASNLib{}, 4.34 pp in \MLFS{}, and 1.12 pp in \SAIL{}. For \Sentinel{}, the performance is similar to that of the baseline.
\RESUB{R2.3}{Further, we manually inspected the mutants killed only by the \texttt{Best} configuration and observed that 70\% of the generated test cases exercise boundary cases missed by the original test suite, which include access to first/last items of arrays and use of boundary values in conditional statement. We can thus conclude that \texttt{Best} helps improve the quality of test suites because, in safety-critical systems, it is desirable to exercise such situations.}
We can thus conclude that \textbf{combining \texttt{Clang}, \texttt{ASAN}, and \texttt{LAF} leads to the best results in \MOTIF}.

\RESUB{R2.5}{We leave the identification of solutions to overcome the mutation score plateau (see Figure~\ref{fig:results:rq1}) to future work; however, it could also be partially due to the presence of equivalent mutants.}

\input{evaluation/table_rq1-comp}

\REV{C2.6}{\emph{Complementarity.} Table~\ref{tab:results:rq1:comp.pos} reports the number of mutants uniquely killed by each MOTIF configuration; they are obtained by subtracting, from the set of mutants killed by one configuration in at least one run, the set of mutants killed by the other configurations in at least one run.
Table~\ref{tab:results:rq1:comp.neg} presents the number of mutants missed by each configuration; they are obtained by subtracting the set of mutants killed by one configuration from the set of mutants killed by all the other configurations. Except for MOTIF with ASAN, MOTIF with UBSAN, and MOTIF with LAF, all the other MOTIF configurations kill a limited number of unique mutants (Table~\ref{tab:results:rq1:comp.pos}). Specifically, MOTIF with UBSAN and MOTIF with LAF are the only configurations killing more than 50 unique mutants; MOTIF with ASAN is the only configuration killing at least one unique mutant in each subject. Looking at the uniquely missed mutants (Table~\ref{tab:results:rq1:comp.neg}), MOTIF with GCC, MOTIF with UBSAN, and MOTIF with MSAN are the configurations leading to the highest number of missed mutants; indeed, each of them leads to more than 20 uniquely missed mutants, the worst case (505 uniquely missed mutants) being MOTIF with UBSAN.}

\REV{C2.6}{To summarize, MOTIF with UBSAN is complementary to other configurations (95 unique mutants killed) but can't be used as the only sanitizer (500 missed mutants); engineers may use it in addition to other configurations, if test budget is available. Also, Table~\ref{tab:results:rq1:comp.pos} and~\ref{tab:results:rq1:comp.neg} confirm the effectiveness of the \texttt{Best} configuration. Indeed, the other best performing configuration options (ASAN and LAF) are part of the \textbf{\texttt{Best}} configuration; further, none of the worst configuration options (i.e., GCC, UBSAN, MSAN) are part of \textbf{\texttt{Best}}.}

%% file: evaluation/table_rq1.tex
\begin{table*}[tb]
    \centering
    \caption{Results obtained by  \MOTIF with different fuzzer configurations. Each cell reports the average percentage of mutants killed over ten runs by each configuration after 10,000 seconds and $p$-values based on U-test against the baseline.}
    \label{tab:results:rq1:sub}
\resizebox{\textwidth}{!}{
    \begin{tabular}{c|c|rrrrr}
    \toprule                           
        \multicolumn{1}{c}{
        \parbox{5em}{\centering{Experiment\\group}}
        } 
            & \multicolumn{1}{c}{Options} 
            & \multicolumn{1}{c}{\UTIL{}} 
            & \multicolumn{1}{c}{\ASNLib{}} 
            & \multicolumn{1}{c}{\MLFS{}} 
            & \multicolumn{1}{c}{\SAIL{}} 
            & \multicolumn{1}{c}{\Sentinel{}} \\
       \midrule
       \multirow{2}{*}{\parbox{6em}{\centering{ Compilers\\ \scriptsize(baseline: GCC)}}}
            & \texttt{GCC} 
                & \cellcolor{green!0 }$48.31\%$  \hspace{3.1em} 
                & \cellcolor{green!0 }$85.44\%$  \hspace{3.1em} 
                & \cellcolor{green!0 }$35.52\%$  \hspace{3.1em} 
                & \cellcolor{green!0 }$38.38\%$  \hspace{3.1em} 
                & \cellcolor{green!0 }$83.03\%$  \hspace{3.1em} \\
            & \texttt{Clang} 
                & \cellcolor{green!20}$49.68\%$ {\footnotesize$(0.0009)$}
                & \cellcolor{green!20}$86.57\%$ {\footnotesize$(0.0002)$} 
                & \cellcolor{green!20}$38.43\%$ {\footnotesize$(0.0002)$} 
                & \cellcolor{green!20}$39.04\%$ {\footnotesize$(0.0001)$} 
                & \cellcolor{green!0 }$82.73\%$ {\footnotesize$(0.2641)$}  \\
        \midrule
        \multirow{4}{*}{\parbox{6em}{\centering{Sanitizers \\ \scriptsize(baseline: Clang)}}}
            & \texttt{ASAN} 
                & \cellcolor{green!0 }$49.22\%$ {\footnotesize$(0.1081)$}
                & \cellcolor{green!10}$86.82\%$ {\footnotesize$(0.0200)$} 
                & \cellcolor{  red!20}$38.29\%$ {\footnotesize$(0.0019)$}
                & \cellcolor{green!20}$39.36\%$ {\footnotesize$(0.0048)$}
                & \cellcolor{  red!20}$81.52\%$ {\footnotesize$(0.0007)$} \\
            & \texttt{UBSAN}      
                & \cellcolor{green!20}$51.35\%$ {\footnotesize$(0.0003)$}
                & \cellcolor{  red!20}$84.31\%$ {\footnotesize$(0.0002)$} 
                & \cellcolor{  red!20}$23.78\%$ {\footnotesize$(0.0002)$}
                & \cellcolor{  red!20}$38.67\%$ {\footnotesize$(0.0001)$}
                & \cellcolor{  red!0 }$82.42\%$ {\footnotesize$(0.2781)$}\\
            & \texttt{LSAN} 
                & \cellcolor{  red!20}$47.39\%$ {\footnotesize$(0.0002)$}
                & \cellcolor{  red!20}$85.78\%$ {\footnotesize$(0.0003)$} 
                & \cellcolor{  red!20}$37.60\%$ {\footnotesize$(0.0002)$}
                & \cellcolor{  red!20}$38.62\%$ {\footnotesize$(0.0002)$}
                & \cellcolor{  red!20}$80.81\%$ {\footnotesize$(0.0000)$} \\
            & \texttt{MSAN} 
                & \cellcolor{  red!20}$45.56\%$ {\footnotesize$(0.0002)$}
                & \cellcolor{  red!20}$83.39\%$ {\footnotesize$(0.0002)$} 
                & \cellcolor{  red!20}$34.33\%$ {\footnotesize$(0.0002)$} 
                & \cellcolor{  red!20}$38.30\%$ {\footnotesize$(0.0001)$} 
                & \cellcolor{  red!20} $6.06\%$ {\footnotesize$(0.0000)$} \\
    \midrule
    \multirow{3}{*}{\parbox{6em}{\centering{Coverage\\metrics \\ \scriptsize(baseline: Clang)}}}
        & \texttt{LAF} 
            & \cellcolor{green!0 }$49.79\%$ {\footnotesize$(0.6158)$}
            & \cellcolor{green!20}$88.84\%$ {\footnotesize$(0.0002)$} 
            & \cellcolor{green!20}$40.09\%$ {\footnotesize$(0.0002)$} 
            & \cellcolor{  red!20}$38.73\%$ {\footnotesize$(0.0004)$} 
            & \cellcolor{  red!0 }$82.53\%$ {\footnotesize$(0.3006)$} \\
        & \texttt{NGRAM} 
            & \cellcolor{green!20}$50.85\%$ {\footnotesize$(0.0016)$}
            & \cellcolor{  red!20}$86.27\%$ {\footnotesize$(0.0023)$} 
            & \cellcolor{  red!20}$38.27\%$ {\footnotesize$(0.0045)$} 
            & \cellcolor{  red!0 }$38.97\%$ {\footnotesize$(0.3446)$} 
            & \cellcolor{green!20}$83.64\%$ {\footnotesize$(0.0005)$} \\
        & \texttt{CTX} 
            & \cellcolor{green!20}$50.89\%$ {\footnotesize$(0.0030)$}
            & \cellcolor{  red!0 }$86.49\%$ {\footnotesize$(0.3989)$} 
            & \cellcolor{  red!20}$38.29\%$ {\footnotesize$(0.0070)$} 
            & \cellcolor{  red!0 }$38.98\%$ {\footnotesize$(0.5518)$} 
            & \cellcolor{green!20}$83.74\%$ {\footnotesize$(0.0001)$} \\

    \bottomrule 
    \multicolumn{7}{r}{\scriptsize
        \colorbox{green!20}{n.nn\%} and \colorbox{green!10}{n.nn\%}: significantly outperformed the baseline with $p\leq 0.01$ and $p\leq 0.05$, respectively 
    } \\
    \multicolumn{7}{r}{\scriptsize
        \colorbox{red!20}{n.nn\%} and \colorbox{red!5}{n.nn\%}: significantly underperformed, compared to the baseline, with $p\leq 0.01$ and $p\leq 0.05$, respectively
    } \\    
    \end{tabular}
}
\end{table*}

%% file: evaluation/fig_rq1.tex
\begin{figure}[ht!]
\begin{center}
        \begin{subfigure}[t]{0.48\columnwidth}\centering
		\includegraphics[width=\columnwidth]{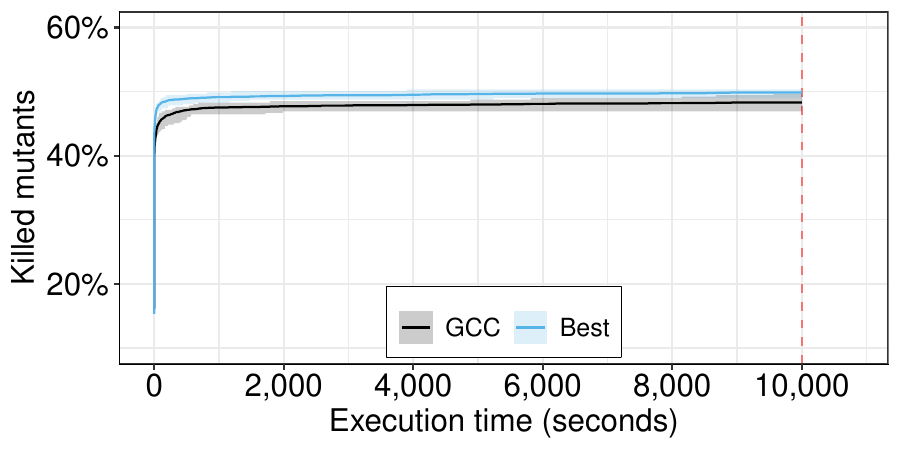}
            \vspace{-0.5em}
		\caption{\UTIL{}}
		\label{fig:results:rq1 LIBU}
	\end{subfigure}
\hfill
        \begin{subfigure}[t]{0.48\columnwidth}\centering
		\includegraphics[width=\columnwidth]{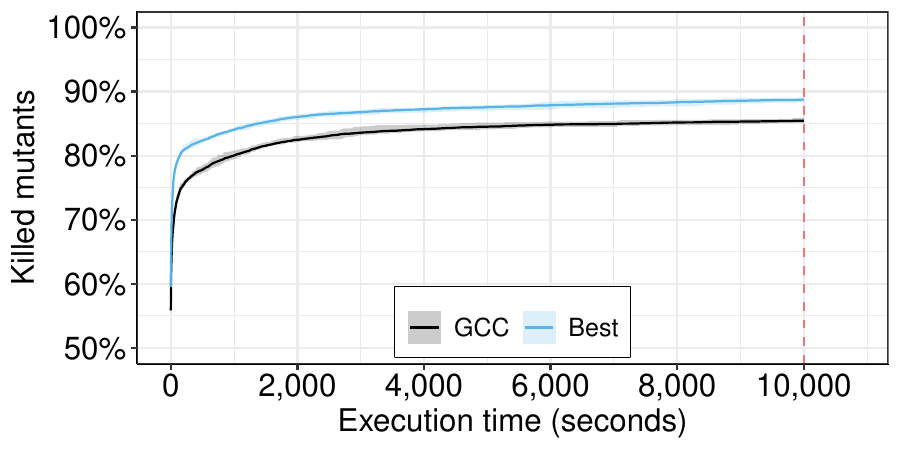}
            \vspace{-0.5em}
		\caption{\ASNLib{}}
		\label{fig:results:rq1 ASN1}
	\end{subfigure}
\bigskip
	\begin{subfigure}[t]{0.48\columnwidth}\centering
		\includegraphics[width=\columnwidth]{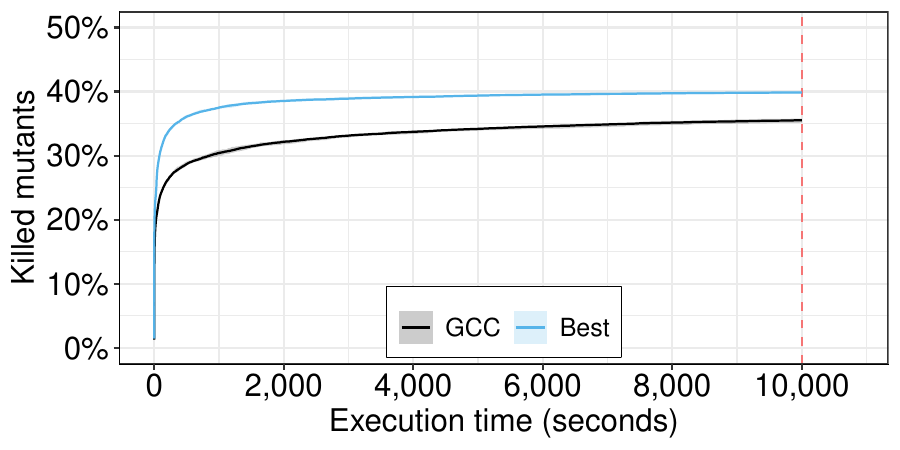}
            \vspace{-0.5em}
		\caption{\MLFS{}}
		\label{fig:results:rq1 MLFS}
	\end{subfigure}
\hfill
	\begin{subfigure}[t]{0.48\columnwidth}\centering
		\includegraphics[width=\columnwidth]{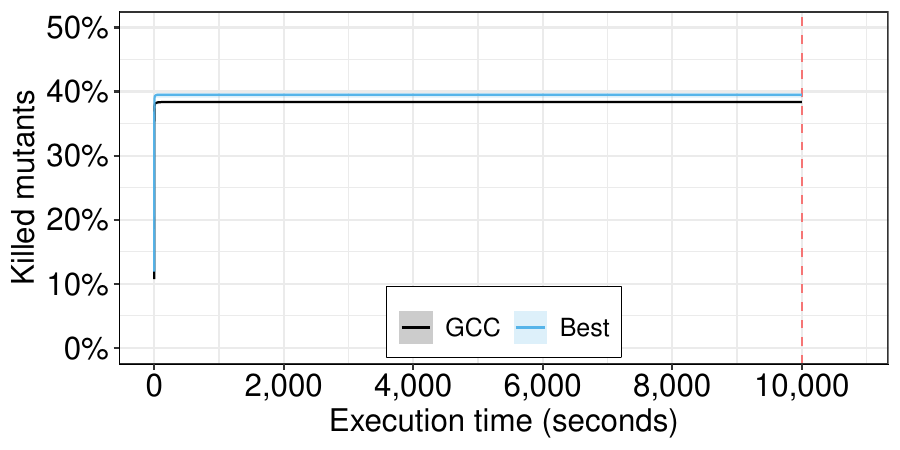}
            \vspace{-0.5em}
		\caption{\SAIL{}}
		\label{fig:results:rq1 ESAIL}
	\end{subfigure}
\bigskip
        \begin{subfigure}[t]{0.48\columnwidth}\centering
		\includegraphics[width=\columnwidth]{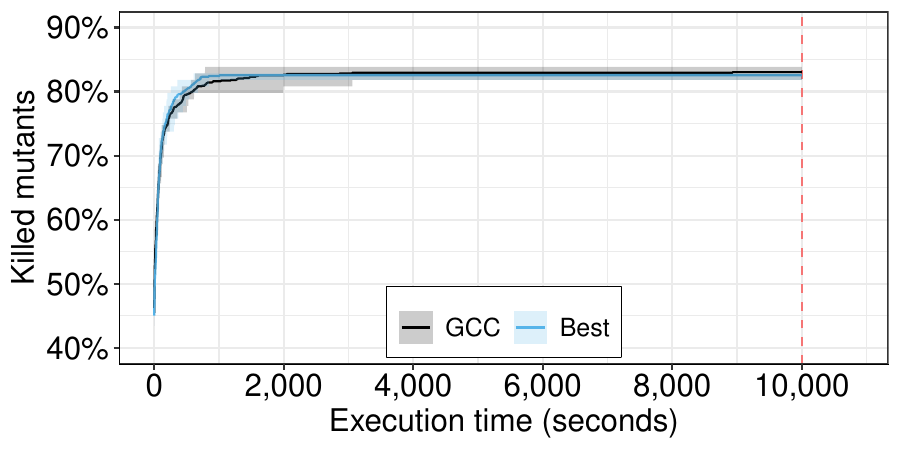}
            \vspace{-0.5em}
		\caption{\Sentinel{}}            
		\label{fig:results:rq1 S5}
	\end{subfigure}
\end{center}
\vspace{-2em}
\caption{Comparison between \MOTIF with \texttt{GCC} and \MOTIF with the \texttt{Best} fuzzer options.}
\label{fig:results:rq1}
\end{figure}

%% file: evaluation/table_rq1-comp.tex
\begin{table*}[tb]
    \centering
    \caption{Uniquely killed mutants per MOTIF configuration.}
    \label{tab:results:rq1:comp.pos}
\resizebox{0.8\textwidth}{!}{
    \begin{tabular}{c|rrrrr}

    \toprule
        	&	\UTIL{}	&	     	\ASNLib{}	   	&	\MLFS{}			&	\SAIL{}	&	\Sentinel{}	\\
    \midrule
        \# of mutants &		437	&		1347		&	3891	     &	581	              &	99	       \\	
    \midrule
        \texttt{GCC}	    &	1 (0.23\%) 	&	 0 (0.00\%) 	&	 7 (0.18\%) 	&	0 (0.00\%) 	&	0 (0.00\%)   \\
        \texttt{CLANG}	&	1 (0.23\%) 	& 	 0 (0.00\%) 	&	 6 (0.15\%) 	&	0 (0.00\%) 	&	0 (0.00\%) \\ 
    \midrule
    \rowcolor{green!20} 
        \texttt{ASAN}	&	1 (0.23\%) 	&	 5 (0.37\%) 	&	 8 (0.21\%) 	&	4 (0.69\%) 	&	1 (1.01\%) \\ 
    \rowcolor{green!20} 
        \texttt{UBSAN}	&	4 (0.92\%) 	&	 0 (0.00\%) 	&	91 (2.34\%)		&   0 (0.00\%) 	&	0 (0.00\%) \\ 
        \texttt{LSAN}	&	0 (0.00\%) 	&	 0 (0.00\%) 	&	 2 (0.05\%) 	&	0 (0.00\%) 	&	0 (0.00\%) \\ 
        \texttt{MSAN}	&	2 (0.46\%) 	&	 0 (0.00\%) 	&	 0 (0.00\%) 	&	1 (0.17\%) 	&	0 (0.00\%) \\ 
    \midrule
    \rowcolor{green!20} 
        \texttt{LAF}	&	2 (0.46\%) 	&	16 (1.19\%)   	&	45 (1.16\%)		&   0 (0.00\%) 	&	0 (0.00\%) \\ 
        \texttt{NGRAM}  &	1 (0.23\%) 	&	 0 (0.00\%) 	&	 4 (0.10\%) 	&	0 (0.00\%) 	&	0 (0.00\%) \\ 
        \texttt{CTX}	&	1 (0.23\%) 	&	 0 (0.00\%) 	&	 3 (0.08\%) 	&	0 (0.00\%) 	&	0 (0.00\%) \\ 
        
    \bottomrule 
    \end{tabular}
}
\end{table*}

\begin{table*}[tb]
    \centering
    \caption{Uniquely missed mutants per MOTIF configuration.}
    \label{tab:results:rq1:comp.neg}
\resizebox{0.8\textwidth}{!}{
    \begin{tabular}{c|rrrrr}
    \toprule
        	&	\UTIL{}	&	     	\ASNLib{}	   	&	\MLFS{}			&	\SAIL{}	&	\Sentinel{}	\\
    \midrule
        \# of mutants &		437	&		1347		&	3891	     &	581	              &	99	       \\	
    \midrule
    \rowcolor{red!20} 
        \texttt{GCC}		&  2 (0.46\%)	&	1 (0.07\%)	&	19 (0.49\%)	&	1 (0.17\%)	&	0 (0.00\%) \\
        \texttt{CLANG}	&  4 (0.92\%)	&	0 (0.00\%)	&	0 (0.00\%)	&	0 (0.00\%)	&	0 (0.00\%) \\
    \midrule
        \texttt{ASAN}	&  0 (0.00\%)	&	0 (0.00\%)	&	0 (0.00\%)	&	0 (0.00\%)	&	0 (0.00\%) \\
    \rowcolor{red!20} 
        \texttt{UBSAN}	&  0 (0.00\%)	&	5 (0.37\%)	&	500 (12.85\%)	&	0 (0.00\%)	&	0 (0.00\%) \\
        \texttt{LSAN}	&  8 (1.83\%)	&	0 (0.00\%)	&	1 (0.03\%)	&	1 (0.17\%)	&	0 (0.00\%) \\
    \rowcolor{red!20} 
        \texttt{MSAN}	&  9 (2.06\%)	&	9 (0.67\%)	&	9 (0.23\%)	&	4 (0.69\%)	&	75 (75.76\%) \\
    \midrule
        \texttt{LAF}		&  0 (0.00\%)	&	0 (0.00\%)	&	0 (0.00\%)	&	0 (0.00\%)	&	0 (0.00\%) \\
        \texttt{NGRAM}	&  0 (0.00\%)	&	1 (0.07\%)	&	0 (0.00\%)	&	0 (0.00\%)	&	0 (0.00\%) \\
        \texttt{CTX}		&  0 (0.00\%)	&	0 (0.00\%)	&	0 (0.00\%)	&	0 (0.00\%)	&	0 (0.00\%) \\
    \bottomrule 
    \end{tabular}
}
\end{table*}

%% file: evaluation/evaluation-RQ2.tex
\subsection{RQ2 - Fuzzing vs Symbolic Execution}

\subsubsection{Design}

We compare fuzzing and symbolic execution in terms of cost-effectiveness.
The effectiveness of an automated mutation testing tool can be measured in terms of the proportion of live mutants killed. Its cost is determined by the time required to kill the mutants; indeed, lengthy test data generation may delay the testing process and increase the usage of computing resources.
Though cost is also driven by the time required to manually inspect test outputs, \MOTIF and \SEMUp should require the same manual inspection time because they invoke the same functions under test and print out the same output values.
Therefore, regarding cost, we focus on execution time and thus compare cost-effectiveness in terms of live mutants killed for different time budgets. 

We could not consider \MLFS to address RQ2 because it works mainly with floating point arguments, which are not supported by KLEE. An old version of KLEE addresses floating point variables but it is not integrated into \SEMU. We could not consider \SAIL{} and \Sentinel{} because of the lack of KLEE support for multi-threading libraries and C++. We therefore focus on \UTIL{} and \ASNLib{}; however, for \UTIL{} we considered only four out of 27 source files, because all the other source files included I/O operations, which are not supported by KLEE/\SEMU, or cannot be compiled into LLVM bitcode. This leads to 1,347 live mutants for \ASNLib{} and 153 for \UTIL{}. For \MOTIF, we considered the \texttt{Best} configuration identified in RQ1.

\subsubsection{Results}

\begin{figure*}[tb]
	\begin{subfigure}[t]{0.49\textwidth}\centering
	    \includegraphics[width=\columnwidth]{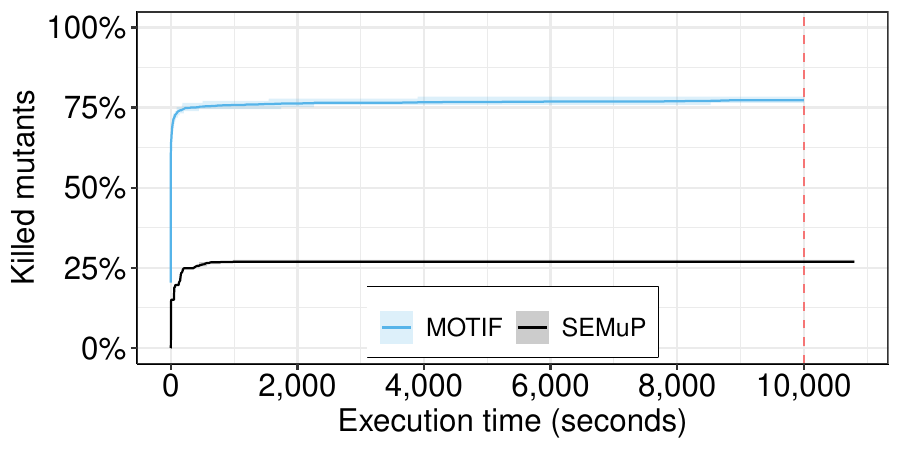}
		\caption{\UTIL{}}
	    \label{fig:rq2 libutil}
	\end{subfigure}
	\hspace{0em}
	\begin{subfigure}[t]{0.49\textwidth}\centering
	    \includegraphics[width=\columnwidth]{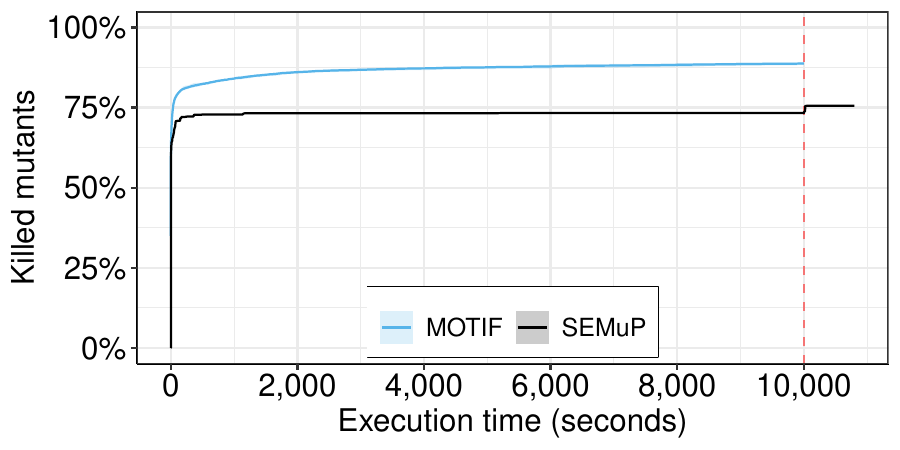}
		\caption{\ASNLib{}}
	    \label{fig:rq2 asn}
	\end{subfigure}
\caption{Percentage of mutants killed by \MOTIF and \SEMUp.}
\label{fig:rq2}
\end{figure*}

Figure~\ref{fig:rq2} depicts the percentage of live mutants killed by \MOTIF and \SEMUp for \UTIL{} (\ref{fig:rq2 libutil}) and \ASNLib{} (\ref{fig:rq2 asn}), respectively.
Each line represents the average percentage from ten runs, with
the shaded area around each line indicating the upper and
lower bounds across those runs. 
The vertical dashed line means the time budget for each experiment.
 At that point, \SEMUp stops exploring paths and generates inputs that satisfy the current path condition, which, sometimes, is sufficient to identify inputs that kill mutants. 
We observe a rapid increase in the number of mutants killed by \SEMUp for the \ASNLib{}, which includes paths with several nested conditions.

The plots show that \MOTIF outperforms \SEMUp. After 10,000 seconds, \MOTIF kills between 117 (76.47\%) and 120 (78.43\%) mutants for \UTIL{} (avg. is 118.20, 77.25\%) and between 1,190 (88.34\%) and 1,200 (89.09\%) for \ASNLib{} (avg. is 1,195.30, 88.74\%). In contrast, \SEMUp kills 41 (26.80\%) to 42 (27.45\%) mutants for \UTIL{} (avg. is 41.2, 26.93\%) and 1,017 (75.50\%) to 1,018 (75.58\%) for \ASNLib{} (avg. is 1,017.8, 75.56\%). 
On average, across the ten runs, \MOTIF kills a percentage of mutants that is 50.33 percentage points (pp) and 13.18 pp higher than \SEMUp's, for \UTIL{} and \ASNLib{}, respectively.

The difference between \MOTIF and \SEMUp is significant at every timestamp, based on a U-test ($p < 0.01$).
For example, after one minute, \MOTIF kills, on average, 110.5 (\UTIL{}) and 1,045.00 (\ASNLib{}) mutants, while \SEMUp kills
29 (\UTIL{}) and 924.6 (\ASNLib{}) mutants. 
For both \UTIL{} and \ASNLib{}, \MOTIF quickly reaches a near plateau; in  \UTIL{}, \MOTIF reaches the plateau quicker because of \UTIL{}'s simple control logic.

Though \MOTIF outperforms \SEMUp, they show some degree of complementarity,  which justifies the integration of hybrid fuzzers in \MOTIF (see Section~\ref{sec:background:fuzzing}).
If we consider the best run of each approach, in the case of \ASNLib{}, \MOTIF kills 257 (19.08\%) mutants not killed by \SEMUp, while \SEMUp kills 75 (5.57\%) mutants not killed by \MOTIF. 
In the case of \UTIL{}, \MOTIF kills an additional 78 (50.98\%) mutants on top of the mutants killed by \SEMUp.  
We manually inspected some of the mutants and noticed that \SEMUp is sometimes better at generating inputs that satisfy narrow, simple constraints. However, such a characteristic is more useful for \ASNLib{}, which mainly performs boundary checks for nested data structures, rather than the utility library. On the other hand, \MOTIF is better when \SEMUp fails to solve complex constraints. For example, for \UTIL{}, \SEMUp could not kill 52 mutants affecting a conditional statement with 24 bitwise operations, 44 mutants affecting a conditional statement with 13 conditions expressed using inequalities, and 5 mutants affecting the size of the buffer used in \texttt{snprintf} statements. \RESUB{R1.f}{Finally, \MOTIF enabled the discovery of four bugs  that were confirmed by developers: two concern missing checks for  out-of-domain numerical inputs, the other two are an integer overflow affecting operations on a numerical data structure with multiple fields. We detected the missing checks by observing that the generated test cases include an out-of-domain input but do not result in an execution error flag being set (likely they were not detected by test cases because the valid domain is underspecified in specifications).  The integer overflow was discovered by observing that the result of a sum of two large numeric items led to a lower number.  The integer underflow can be noticed by observing that the difference between a zero-filled data instance and a data instance with small positive numbers result in large positive numbers.} \SEMUp discovered only three of them.

%% file: evaluation/evaluation-RQ3.tex
\subsection{RQ3 - Seeding effectiveness}

\subsubsection{Design}

To discuss how \MOTIF's seeds contribute to mutation testing results, we focus on the proportion of mutants killed with seed inputs in the experiments performed to address RQ1 with the \MOTIF's \texttt{Best} configuration.

\subsubsection{Results}

\MOTIF's seed inputs contributed to killing mutants as follows: 22.2 (5.08\%) for \UTIL{}, 304 (22.57\%) for \ASNLib{}, 81 (2.08\%) for \MLFS{}, 69 (11.88\%) for \SAIL{}, and 39 (39.39\%) for \Sentinel{}. However, although in all the subjects except \UTIL{} seed inputs killed the same mutants across the ten runs, in \UTIL{} the number of killed mutants varied between 22 and 23 because of a non-deterministic function that computes the time difference before and after invoking the \texttt{sleep()} function.

The percentage of mutants killed by seed inputs largely depends on the nature of the functions under test. For mutants in \MLFS and \UTIL{}, such percentages are low because they mainly alter mathematical operations whose mutants are killed with inputs satisfying complex constraints. 
For \ASNLib{}, \SAIL{}, and \Sentinel{}, the proportion of mutants killed is higher because the mutants modified lines that affect the output observed with any input value.
Please note that seed inputs do not introduce bias in RQ2 results since \SEMUp kills most of the mutants killed by seed inputs ($267/304$ for \ASNLib{} and $1/1$ for \UTIL{}).

Concluding, although the selected seed inputs help kill mutants, the contribution of the fuzzing process is significant with, at the very minimum (\Sentinel{}), \REV{C1.6}{52.27\% (i.e., $100\%-\frac{39.39\%}{82.53\%}$)} of the killed mutants being killed by fuzzing.

%% file: evaluation/evaluation-RQ4.tex
\subsection{RQ4 - Applying Hybrid-fuzzing}

\subsubsection{Design}
We compare \MOTIF with \MOTIF-\textit{Hybrid}, our \MOTIF extension integrating \texttt{AFL++} with \texttt{SymCC} to leverage hybrid fuzzing for mutation testing (see Section~\ref{sec:evaluation:setup}).
To ensure a fair comparison, we applied the \texttt{Best} fuzzing configuration identified in RQ1 to both approaches.
As for RQ2, we compare the two approaches in terms of cost-effectiveness, by reporting on the MS  and the time taken to kill mutants.

We considered four subjects: \UTIL{}, \ASNLib{}, \MLFS{}, and \SAIL{}. We excluded \Sentinel{} because \texttt{SymCC} cannot successfully compile C++ code relying on the \texttt{STL} library.
Also, since \texttt{SymCC} fails to correctly compile code chunks with increment and decrement operators for \texttt{boolean}, we excluded two \SAIL{} mutants.
Finally, we executed the two approaches ten times on each subject for 10,000 seconds.

\begin{figure*}[b]
	\begin{subfigure}[t]{0.49\textwidth}\centering
	    \includegraphics[width=\columnwidth]{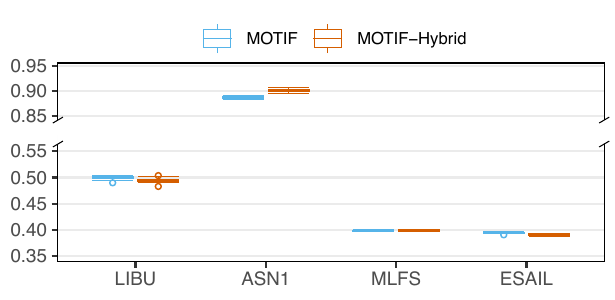}
		\caption{Mutation score}
	    \label{fig:results:rq4 ms}
	\end{subfigure}
	\hspace{0em}
	\begin{subfigure}[t]{0.49\textwidth}\centering
	    \includegraphics[width=\columnwidth]{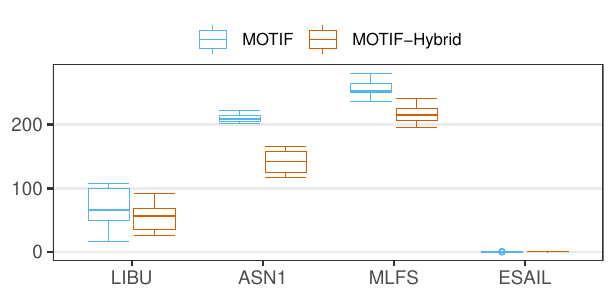}
		\caption{Kill time (seconds)}
	    \label{fig:results:rq4 time}
	\end{subfigure}
\caption{Results observed when applying \MOTIF with \texttt{AFL++} (\emph{\MOTIF}) and \MOTIF with \texttt{SymCC} (\emph{\MOTIF-Hybrid}).}
\label{fig:results:rq4}
\end{figure*}

\subsubsection{Results}
Figure~\ref{fig:results:rq4} presents boxplots 
providing the distributions of MS~(\ref{fig:results:rq4 ms}) after 10,000 seconds and the time taken to kill mutants (\ref{fig:results:rq4 time}), across the ten experimental runs. 

\MOTIF-\textit{Hybrid} increases the percentage of killed mutants in \ASNLib{} by up to 1.37 pp (18.5 mutants) on average, while the performance differences for the other subjects are not statistically significant.

As for the time taken to kill mutants, we report that the mutants in \ASNLib{} and \MLFS were killed significantly faster with \MOTIF-\textit{Hybrid}. 
\REV{C1.3}{When compared with \MOTIF, the median kill time per mutant of \MOTIF-\textit{Hybrid} is reduced by 31.95\% for \ASNLib{} (from 208.66 to 142.00 seconds) and by 15.20\% for \MLFS{} (from 253.08 to 214.60 seconds).}
In \UTIL{} and \SAIL{}, the difference is not significant, however, with the median observed for \MOTIF-\textit{Hybrid} with \UTIL being lower \REV{C1.3}{by 13.65\%} (57.13 seconds with \MOTIF-\textit{Hybrid} vs. 66.13 seconds with \MOTIF).

\input{evaluation/table_rq4}

Additionally, we manually investigated the complementarity between \MOTIF-\textit{Hybrid}, \MOTIF, and \SEMUp in \ASNLib{}, which, in RQ2, showed that \APPR and \SEMUp are complementary. 
Table~\ref{tab:results:rq4:complementarity} shows the number of killed and live mutants observed in the best run of each approach.
\MOTIF-\textit{Hybrid} killed 267 mutants (19.82\%) that were not killed by \SEMUp.
However, \MOTIF-\textit{Hybrid} killed only 12 of the 75 mutants killed by \SEMUp but not by \MOTIF; the fact that \texttt{SymCC} cannot kill all the mutants killed by \SEMUp shows that \texttt{SymCC} presents some limitations compared to \SEMUp (e.g., it may need to augment the set of inputs executed symbolically).
However, if we subtract
the mutants missed by \MOTIF-\textit{Hybrid} from the additional mutants killed by \MOTIF-\textit{Hybrid} compared to \SEMUp and \MOTIF, we observe that \MOTIF-\textit{Hybrid} still kills 204 more mutants than \SEMUp and 22 more mutants than \MOTIF, thus remaining the most effective choice.

Concluding, \textbf{\MOTIF-\textit{Hybrid} is the best approach for mutation testing since it is faster at killing mutants and has a higher MS in half of the subjects, while getting an equivalent MS with the others, when technical limitations do not prevent its adoption (e.g., in \texttt{C++} projects like \Sentinel{})}. \RESUB{??}{\textbf{\MOTIF-\textit{Hybrid} could therefore be a practical solution for large projects resulting in a significant number of mutants.}} However, the setup required for \MOTIF-\textit{Hybrid} is more complex than the one required for \MOTIF; indeed, 
it is necessary to set up additional configurations for the AFL fuzzer and to compile the fuzzing drivers twice, with the AFL compiler and with SYM-CC.
Consequently, \MOTIF-\textit{Hybrid} is less likely to be adopted in practice, in favor of \MOTIF.

%% file: evaluation/table_rq4.tex
\begin{table}[t!]
\renewcommand{\arraystretch}{1.3}
\renewcommand{\tabcolsep}{4pt}
\small
\centering
\caption{Complementarity between \MOTIF-\textit{Hybrid}, \MOTIF, and \SEMUp in \ASNLib{}: number of mutants killed and not killed by the selected approaches, for their best execution.}
\label{tab:results:rq4:complementarity}
\begin{tabular}{ccrrcrr}
    \toprule               
        & & \multicolumn{2}{c}{\MOTIF-\textit{Hybrid}} & & \multicolumn{2}{c}{\SEMUp} \\ 
        \cline{3-4} \cline{6-7}
        & & KILLED  &	LIVE & & KILLED  &	LIVE \\ 
    \midrule
        \multirow{2}{*}{\centering{\SEMUp}}
        & KILLED &    955  &   63 & & -& -\\
	&   LIVE &    267  &   62 & & -& - \\
    \midrule
        \multirow{2}{*}{\centering{\MOTIF}}
        & KILLED &  1188 &   12  & &   943 &	 257\\
	& LIVE   &    34 &  113  & &    75 &	  72\\
    \bottomrule
\end{tabular}
\end{table}

%% file: evaluation/evaluation-RQ5.tex
\subsection{RQ5 - Reusing inputs}

\subsubsection{Design}
\label{RQ5:design}
RQ5 aims to assess the tradeoff between mutation score improvement and increased testing time observed when reusing inputs killing mutants to kill other mutants not successfully killed by \MOTIF.
For each mutated function with at least one mutant killed by \MOTIF, we rely on the mutant-killing inputs generated by \MOTIF to test the mutants not successfully killed by \MOTIF. Precisely, we collect the fuzzed files that kill mutants in a function and, after excluding duplicates, provide them as inputs to the fuzzing driver for a live mutant of the same function. We also apply the \MOTIF post-processing step to avoid false positives (see Step 4 in Section~\ref{sec:stepFour}).

We considered all the five subjects used for RQ1, and the mutant-killing inputs obtained with the \texttt{Best} configuration. 
To account for randomness, we apply the approach to the results of all the ten runs considered for RQ1.

To discuss effectiveness, we compare the MS obtained by reusing inputs with the MS obtained by \MOTIF after 10,000 seconds. %

As for cost, we measure, for each mutant, the time taken to execute fuzzing drivers with all inputs and the time taken until the mutant is killed. Since the order of inputs may affect execution time, we repeat our experiment 50 times, after shuffling each input set, \REV{2.14}{for each of the 10 MOTIF runs for the \texttt{Best} configuration considered in RQ1. Our setting leads to 500 different results for execution time, in total. By construction, all the 50 repetitions lead to the same mutation score (the final mutation score is not based on the input processing order); consequently, we have 10 distinct mutation score results, not 500.}

\subsubsection{Results}
Table~\ref{tab:results:rq5:live} presents our results. Column \texttt{target mutants} reports the number of mutants that belong to the functions that have at least one killed mutant. Column \texttt{FUZZED} reports the MS obtained by \MOTIF on average over 10 runs. Column \texttt{REUSED} reports the MS obtained after reusing inputs to kill additional mutants. Column \texttt{DIFF} shows the differences between \texttt{FUZZED} and \texttt{REUSED}, and reports on the significance of the difference (we highlight significant improvements).
The next three sets of columns report statistics (mean, min, and max) on the number of inputs available for reuse across target functions, the time required to test each mutant with the reused inputs, and the time taken to kill a mutant. \REV{2.14}{Regarding the execution time, we report statistics based on the 500 execution results obtained as described in Section~\ref{RQ5:design}.}

\REV{C1.3}{The experiment results show that, in the best case (i.e., \UTIL{}), the MS improves by 3.79 pp and, in the worst case (i.e., \SAIL{}), it improves by 0.12 pp.  The average improvement across the five subjects is 1.40 pp. Since in most of the subjects the improvement is above 1 pp, we conclude that input reuse is beneficial for \MOTIF.}
The time taken for reusing inputs varies, taking up to 15 minutes (900 seconds) depending on several factors, including the number of inputs, software size, and the type of mutants.
For instance, \SAIL{} has approximately 70\% of its mutants within a single function, 
which leads to a large number of additional executions when reusing inputs. In the case of \Sentinel{}, the larger size causes a longer execution time for fuzzing drivers compared to other subjects, but the maximum execution time is lower due to the number of inputs collected. The execution time is also affected by the type of mutants. Specifically, some mutants took longer than others because they modified stopping conditions in program loops.

\input{evaluation/table_rq5}

Since, in the worst case, input reuse takes 855 seconds, and given that Figure~\ref{fig:results:rq1} shows that after 9000 seconds \MOTIF has already reached the plateau, we can conclude that with a test budget of 10,000 seconds per mutant, it is convenient to rely on input reuse. For lower test budgets input reuse may provide benefits only for a subset of subjects. However, based on our results, up to 5000 seconds of test budget, input reuse is likely to be beneficial; indeed, in all our subjects, after 4150 seconds, \MOTIF has killed at most 1 pp mutants less than at 10,000 seconds. Concluding, \textbf{we suggest executing \MOTIF with a budget of 5000 seconds per mutant, and dedicate the last 900 seconds to input reuse}.%

\ENDUPDATED

%% file: evaluation/table_rq5.tex
\begin{table*}[t]
\renewcommand{\arraystretch}{1.3}
\renewcommand{\tabcolsep}{4pt}
\small
\centering
\caption{Comparison \MOTIF with \texttt{Best} set of fuzzing options and results after reusing inputs for five selected subjects.}
\label{tab:results:rq5:live}
\resizebox{\textwidth}{!}{
    \begin{tabular}{crp{1pt}rrrp{0.1em}rrrp{0.1em}rrrp{0.1em}rrr}
    \toprule               
        \multirow{2}{*}{Subject} 
        & \multirow{2}{*}{\parbox{3em}{\centering{Target \\ mutants}}} 
        & \parbox{1em}
        & \multicolumn{3}{c}{\parbox{10em}{\centering{{MS} \\ (Avg. of 10 runs)}}} 
        & \parbox{1em}
        & \multicolumn{3}{c}{\parbox{9em}{\centering{\# of inputs for \\ functions (10 runs)}}} 
        & \parbox{1em}
        & \multicolumn{3}{c}{\parbox{8em}{\centering{Reuse time \\ ({10$\times$50} runs, seconds)}}} 
        & \parbox{1em}
        & \multicolumn{3}{c}{\parbox{8em}{\centering{Kill time \\ ({10$\times$50} runs, seconds)}}} \\
        \cmidrule{4-6}\cmidrule{8-10}\cmidrule{12-14}\cmidrule{16-18}
    
           &
           && \multicolumn{1}{c}{FUZZED}  
           & \multicolumn{1}{c}{REUSED} 
           & \multicolumn{1}{c}{DIFF}  
           && \multicolumn{1}{c}{Mean}
           & \multicolumn{1}{c}{Min}
           & \multicolumn{1}{c}{Max}
           && \multicolumn{1}{c}{Mean}
           & \multicolumn{1}{c}{Min}
           & \multicolumn{1}{c}{Max}
           && \multicolumn{1}{c}{Mean}
           & \multicolumn{1}{c}{Min}
           & \multicolumn{1}{c}{Max} \\ 
    \midrule

        \UTIL{}   &  325   &&     61.78\%  & 62.71\% & \cellcolor{green!20}0.92 pp (0.0048)
            && 299.49 & 1 & 3166   &&   5.72 & 0.01 & 264.84 &&      0.21 & 0.02 & 5.02 \\
        \ASNLib{} &   762   &&   80.09\%  & 83.88\% & \cellcolor{green!20}3.79 pp (0.0002)
            && 143.69 & 4 &  779 &&   9.13 & 0.10 & 854.79 &&      0.12 & 0.01 & 1.06 \\
        \MLFS{}   & 3623   &&   42.50\%  & 43.57\% & \cellcolor{green!20}1.08 pp (0.0002)
            && 125.65 & 1 & 1267 &&  12.51 & 0.01 & 353.51 &&      0.11 & 0.01 & 4.14 \\
        \SAIL{}   &   418   &&    54.90\%  & 55.02\% & \cellcolor{green!0}0.12 pp (0.2515)
            && 246.78 & 1 & 1398&& 484.40 & 0.23 & 835.10 &&      0.24 & 0.14 & 1.06 \\
        \Sentinel{} &    28 &&    88.21\%  & 89.29\% &  \cellcolor{green!0} 1.07 pp (0.0767)
            &&  11.33 & 7 &   28 &&   1.76 & 0.00 &   6.74 &&      0.10 & 0.08 & 0.15 \\

    \bottomrule 
    \multicolumn{18}{r}{\scriptsize
        \colorbox{green!20}{n.nn pp}: outperformed the baseline with $p\leq 0.01$ 
    } \\
    \end{tabular}
}
\end{table*}

%% file: conclusion.tex
\section{Conclusion}
\label{sec:conclusion}

We propose \MOTIF, an approach that leverages fuzzing to automatically generate test data for mutation testing of embedded software deployed in cyber-physical systems (CPS). It aims to overcome the limitations of SOTA approaches, which rely on symbolic execution and cannot easily be applied in many contexts, especially CPS ones. 

\MOTIF is implemented through a pipeline that generates a test driver that processes the input data generated by the fuzzer, provides
such input data 
to the original and mutated versions of a function under test, and determines when the outputs generated by the two functions differ (i.e., the mutant is killed). By monitoring the coverage achieved when executing the original and mutated functions, the fuzzer identifies inputs leading to different behaviors across these functions and, consequently, is driven towards the identification of inputs that kill the mutant.

We performed an empirical evaluation with embedded software deployed on satellites currently in orbit. 
We empirically determined the fuzzer configurations leading to the best mutation testing results, which consists of relying on the Clang compiler with address sanitization and LAF coverage optimization. In our subjects, such configuration enables killing between 40\% and 83\% of the mutants.
Further, although our seeding strategy contributes to quickly killing mutants, most of the mutants (between 60\% and 97\%) are killed thanks to the fuzz testing process.
We compared \MOTIF with a SOTA approach based on symbolic execution, which showed that the percentage of mutants killed by \MOTIF is higher than the SOTA approach by 13 and 50 percentage points in our two case studies where symbolic execution is applicable. Our results therefore clearly show that fuzzing should be adopted as the preferred method to use to perform mutation testing.
However, we also demonstrated that hybrid-fuzzing, which integrates fuzzing and symbolic execution, leads to  slightly increasing  the percentage of killed mutants (up to 1.37 pp).

%% file: ack.tex
\section*{Acknowledgment and Declarations}

\noindent\textbf{Funding:} This research was supported by ESA via a GSTP element contract (RFQ/3-17554/21/NL/AS/kkIMPROVE), by the NSERC Discovery and Canada Research Chair programs, and by 2025 Research Grant from Kangwon National University. The experiments presented in this paper were carried out using the HPC facilities of the University of Luxembourg (see \url{http://hpc.uni.lu}).

\noindent\textbf{Ethical approval:} This article does not contain any studies with human participants or animals performed by any of the authors.

\noindent\textbf{Informed consent:} Not applicable.

\noindent\textbf{Author contributions:}
All authors contributed to the study conception and design. Material preparation, data collection, and analysis were performed by Jaekwon Lee, Fabrizio Pastore, and Lionel Briand. The tool implementation was performed by Jaekwon Lee. The first draft of the manuscript was written by Fabrizio Pastore, and all authors reviewed and changed previous versions of the manuscript. All authors read and approved the final manuscript.
The authors would like to thank Oscar Cornejo and Enrico Vigan{\`{o}} for having contributed to the preliminary versions of \MOTIF. 

\noindent\textbf{Data availability:} 
Part of data collected and used for the experiments conducted in this study—including the datasets, software, and scripts for generating visualizations—are publicly available at the following repository: \url{https://figshare.com/s/5a9a1fa723c374f5d0fd}.
However, data related to industrial partners is not publicly accessible.

\noindent\textbf{Conflict of interest:} The authors have no relevant financial or non-financial interests to disclose.

\noindent\textbf{Clinical trial number:} not applicable.